\documentclass[fleqn,usenatbib]{mnras}

\usepackage{newtxtext,newtxmath}
\usepackage{siunitx} 
\usepackage{graphicx}
\usepackage{subcaption}
\usepackage{xspace}     % Appropriate space for commans
\DeclareSIUnit{\solarmass}{\ensuremath{M_\odot}\xspace}

\usepackage[T1]{fontenc}
\usepackage{array}
\usepackage{cancel}
\DeclareRobustCommand{\VAN}[3]{#2}
\let\VANthebibliography\thebibliography
\def\thebibliography{\DeclareRobustCommand{\VAN}[3]{##3}\VANthebibliography}

\usepackage{graphicx}	% Including figure files
\usepackage{amsmath}	% Advanced maths commands

\usepackage{xspace}     % Appropriate space for commans
\usepackage{booktabs}

\newcommand{\nulupi}[1]{$\nu ^2$ lupi}
\newcommand{\numax}{\ensuremath{\nu_\text{max}}\xspace}
\newcommand{\dnu}{\ensuremath{\Delta\nu}\xspace}
\newcommand{\deltanu}{\dnu}
\newcommand{\teff}{\ensuremath{T_{\textup{eff}}}\xspace}
\newcommand{\feh}{\ensuremath{[\text{Fe/H}]}\xspace}
\newcommand{\meh}{\ensuremath{[\text{M/H}]}\xspace}
\newcommand{\alphafe}{\ensuremath{[\text{$\alpha$/Fe}]}\xspace}

\newcommand{\ironmassfrac}{\ensuremath{\text{\%Fe}}\xspace}
\newcommand{\starironmassfrac}{\ensuremath{F_{\text{iron}}^{\text{star}}}\xspace}

\newcommand{\planetradii}{\ensuremath{R_{\rm{p}}}\xspace}
\newcommand{\earthradius}{\ensuremath{R_{\oplus}}\xspace}

\newcommand{\qflag}[1]{\texttt{#1}\xspace}

\let\apj=\aj

\newcommand{\psj}{PSJ}

\usepackage{adjustbox}%
\newcommand{\basta}{\texttt{BASTA}\xspace}
\newcommand{\stellarmass}{\ensuremath{M_{\star}}\xspace}
\newcommand{\sweetcat}{SWEET-Cat\xspace}
\newcommand{\linmix}{\texttt{linmix}\xspace}
\newcommand{\Garstec}{\textsc{Garstec}\xspace}\newcommand{\adipls}{\textsc{adipls}\xspace}

\title[A Link between Rocky Planet Composition and Stellar Age]{A Link between Rocky Planet Composition and Stellar Age}

\author[A. Weeks et al.]
{Angharad Weeks,$^{1,2}$\thanks{E-mail: angharad.weeks.20@ucl.ac.uk}
Vincent Van Eylen$^{1}$,
Daniel Huber,$^{2,3}$
Daisuke Kawata,$^{1}$
Amalie Stokholm,$^{4,5}$ \newauthor 
Victor Aguirre B{\o}rsen-Koch,$^{6}$ 
Paola Pinilla,$^{1}$
Jakob Lysgaard R{\o}rsted, $^{5,7}$
Mark Lykke Winther, $^{5}$
Travis Berger $^{8}$
\\
$^{1}$Mullard Space Science Laboratory, University College London, Dorking, Surrey, RH5 6NT\\
$^{2}$Institute for Astronomy, University of Hawai‘i, 2680 Woodlawn Drive, Honolulu, HI 96822, USA \\
$^{3}$Sydney Institute for Astronomy, School of Physics, University of Sydney NSW 2006, Australia\\
$^{4}$School of Physics and Astronomy, University of Birmingham, Edgbaston, Birmingham, B15 2TT, United Kingdom \\
$^{5}$Stellar Astrophysics Centre, Department of Physics and Astronomy, Aarhus University, Ny Munkegade 120, 8000, Aarhus, Denmark \\
$^{6}$DARK, Niels Bohr Institute, University of Copenhagen, Jagtvej 128, 2200, Copenhagen, Denmark \\
$^{7}$Aarhus Space Centre (SpaCe), Department of Physics and Astronomy, Aarhus University, Denmark \\
$^{8}$Space Telescope Science Institute, 3700 San Martin Drive, Baltimore, MD 21218, USA \\
}

\date{Accepted XXX. Received YYY; in original form ZZZ}

\pubyear{2015}

\begin{document}
\label{firstpage}
\pagerange{\pageref{firstpage}--\pageref{lastpage}}
\maketitle
\setlength{\extrarowheight}{8pt}
% Abstract of the paper
\begin{abstract}
Interior compositions are key for our understanding of Earth-like exoplanets. The composition of the core can influence the presence of a magnetic dynamo and the strength of gravity on the planetary surface, both of which heavily impact thermal and possible biological processes and thus the habitability for life and its evolution on the planet. However, detailed measurements of the planetary interiors are extremely challenging for small exoplanets, and existing data suggest a wide diversity in planet compositions.
Hitherto, only certain photospheric chemical abundances of the host stars have been considered as tracers to explain the diversity of exoplanet compositions.
Here we present a homogeneous analysis of stars hosting rocky exoplanets, with ages between 2 and 14 Gyr, revealing a correlation between rocky exoplanet compositions and the ages of the planetary systems. Denser rocky planets are found around younger stars. This suggests that the compositional diversity of rocky exoplanets can be linked to the ages of their host stars.
We interpret this to be a result of chemical evolution of stars in the Milky Way, which modifies the material out of which stars and planets form.
The results imply that rocky planets which form today, at similar galactocentric radii, may have different formation conditions, and thus different properties than planets which formed several billion years ago, such as the Earth.
\end{abstract}

% Select between one and six entries from the list of approved keywords.
% Don't make up new ones.

%%%%%%%%%%%%%%%%%%%%%%%%%%%%%%%%%%%%%%%%%%%%%%%%%%

%%%%%%%%%%%%%%%%% BODY OF PAPER %%%%%%%%%%%%%%%%%%

\section{Introduction}
The internal composition of exoplanets is a key characteristic to understand their diversity and formation. 
The interior of small, Earth-like planets can be approximated as having a dense metallic core, a lower and upper silicate mantle, and a thin layer of ice or liquid water \citep{Valencia2007}. 
The amount of elements such as iron (Fe), silicon (Si) or magnesium (Mg) plays an important role in influencing the extent of each of these layers \citep{Sotin2007}. 
In the solar system, the relative abundances of Fe, Si and Mg are similar for the Sun and Earth, Mars and Venus \citep{McDOnough2021}.
The properties of the stars around which planets form are expected to influence planetary interiors and determine the planetary core size \citep{Dorn2015}. These properties in turn directly influence characteristics such as the presence or absence of a magnetic field, which is driven by the dynamo of the molten iron core \citep{Taylor1963}, the importance of outgassing to create a secondary atmosphere through plate tectonics \citep{Muller2022}, the presence of a carbon cycle which may affect habitability \citep{Alibert2015} and the strength of gravity on the planet as its size changes for a given mass \citep{Yang2019}.
\par
Understanding the interior compositions of rocky exoplanets is challenging. A limited number of small planets with known atmospheres have been observed with space-based spectroscopy \citep[e.g.,][]{kreidberg2014,mikalevans2023,kempton2023}. Several small rocky planets have been observed by JWST, but thus far, no atmospheres have been detected \citep[e.g.][]{Lim2023,Zieba2021}. 

When such atmospheric measurements are not possible or an atmosphere is not present, the determination of exoplanet compositions relies on measurements of a planet's mass and radius, which then yields its bulk density. These measurements can then be used to derive parametric mass-radius relationships \citep[e.g.,][]{Weiss+Marcy2014,Wolfgang2016,chen17,Bashi2017,Otegi2020} or infer compositions through theoretical mass-radius tracks that solve the equation of states of the varying elemental constituents \citep[e.g.,][]{Seager2007, Sotin2007,Zeng+seager2008,Zeng2016}.

\par
The interpretation of mass-radius measurements for planets that have both a core and an atmosphere, as for planets with a significant hydrogen-helium (H-He) envelope, are often highly degenerate \citep{Lopez+Fortney2014,Kite2020,selsis2007,Bean2021}.
Planets without a thick atmosphere are more suited for studies of planetary core compositions. As super-Earths are expected to be stripped cores, their interior composition models are generally more uniquely defined
\citep{Rogers2015,Valencia2007}. This makes it feasible to infer the extent of the dense Fe-rich core for such planets through an estimate of the planet mean density.
%
%\par
%One way to select planets that have no significant Hydrogen-Helium atmosphere 
One way to select these planets is by exploiting the so-called radius valley, a dearth in occurrence of planets with radii of about 1.8~$R_{\oplus}$ \citep{KCSradiusvalley2017}. The valley is a likely consequence of atmospheric mass loss \citep{owenandwu,Gupta_Schliting_2019}
%of photo-evaporation or core-powered mass loss (e.g., \citep{owenandwu} \citep{Gupta_Schliting_2019}), 
and separates planets with a significant H-He atmosphere (sub-Neptunes) from those that have lost their atmosphere (super-Earths). Super-Earths sit below the valley at sizes smaller than about two Earth radii, though the exact location of the valley is dependent on orbital period \citep{vaneylen2018} and stellar type \citep[e.g.,][]{KCSradiusvalley2018,Ho+Vanylen2022}. 
%Because super-Earths are expected to be stripped cores, they are ideally suited to study planetary core composition since their interior composition models are generally more uniquely defined \citep{Rogers2015,Valencia2007}. This makes it feasible to infer the extent of the dense Fe-rich core for such planets through an estimate of the planet mean density. 
\par
To measure the mean density of a planet, both its radius and its mass need to be estimated. For the majority of known exoplanets, radii and masses are inferred by combining transit and radial velocity observations \citep[e.g.,][]{lam2021,vaneylen2021}. Both quantities are measured relative to the radii and masses of the host star, which are often determined heterogeneously. Individual planet discovery and characterisation papers use different atmospheric parameters, stellar spectra models, stellar evolution models, and fitting codes, in order to calculate stellar parameters.
\par
Here we homogenise the parameters for a sample of stars with planets that have masses and radii measured by the transit and radial velocity methods, and we use these new parameters to refine the properties of the planets. 
We then link the properties of these planets to the properties of the host stars, and discuss an observed correlation between stellar age and rocky planet composition, which we interpret in the context of galactic chemical evolution. 

In Section~\ref{sec:stellar}, we outline how our sample was selected and how we homogeneously characterised the host stars. We also validate the stellar parameters. In Section~\ref{sec:planet}, we describe how we use the homogenised stellar parameters to determine new planet parameters. In Section~\ref{sec:agecomposition}, we link properties of the planets to their star, and outline an observed correlation between stellar age and rocky planet composition. We discuss this finding in the context of galactic chemical evolution in Section~\ref{sec:discussion}. In Section~\ref{sec:conclusions}, we present our conclusions.

\section{Homogeneous host star characterisation}
\label{sec:stellar}

\subsection{Sample selection}

We start by selecting all confirmed planets with $\planetradii <  4 \earthradius$ from the NASA Exoplanet 
Archive\footnote{https://exoplanetarchive.ipac.caltech.edu/index.html, last accessed 15/11/2023} \citep{Akeson2013}. We also require that the planets have both radius and mass measurements from transit and radial velocity observations. This means that values in both \qflag{rv\_flag} and \qflag{trans\_flag} must be 1. %The latter point results in a sample of relatively bright stars (with mean $M_G = 10.4$).
%many of which have multiple high-resolution spectra available, and enables us to focus on the small planet radius valley and mass - radius relations for small exoplanets.
\par

\par
The DR2 Gaia ID in the NASA exoplanet Archive is matched to the corresponding DR3 Gaia ID using the \verb+`DR2_neighbours'+ table in the Gaia archive, following the algorithm in \citep{Marrese2019}. Using Python modules \texttt{astropy} \citep{astropy} and \texttt{astroquery} \citep{astroquery}, the sample is then cross-matched to the Gaia DR3 Astrophysical parameters supplementary materials  \citep{Aprams23}. Here, stars with a renormalised unit-weighted error (\qflag{RUWE}) of $>1.4$, or parallax over error $<5$ are removed as these targets are likely to be either binaries \citep{belokurov20} or have large uncertainties on their heliocentric distance, respectively. \cite{Lester2021} note that there is a decreased efficiency of planet detection around binary stars, for planets with radii less than 2~R$_{\oplus}$. We note that 55 Cnc is a known long-period binary system \citep{Bourrier2018}, however its Earth-like planet, 55 Cnc e, has been observed and analysed in many planet characterisation and follow-up papers over the last 15 years \citep[see e.g.,][]{Dai2019, Demory2016, Winn2011}). Kepler-20 and Kepler-21 have potentially detected comparisons from more in-depth astrometry \cite{Mugrauer2019} and imaging \cite{Ginski2016}, respectively. However, our stellar parameters agree well with those in the literature, as these systems are extremely well-characterised with asteroseismology. We require that each target has a value reported for \qflag{teff\_gspspec}, \qflag{mh\_gspspec} and their associated upper and lower confidence intervals. Further, we remove stars with \qflag{teff\_gspspec}$< 5000$~K, for which model ages cannot be reliably measured \citep{Berger2020}.
Some stars have reported non-zero quality flags in the first 13 flag parameters, described in \citep{Recio-Blanco2022a}. We confirm that no flag is greater than a value of 2 for these first 13 flags.

\par
Following these selection criteria, we are left with a sample of 89 stars hosting 116 small planets with measured masses and radii. They all have \teff and \meh from GSP-Spec, Gaia $G$, $RP$ and $BP$ photometry, and parallax, $\varpi$, from Gaia DR3, in addition to radial velocity K amplitudes and transit-derived $\frac{R_p}{R_\star}$ values from the NASA Exoplanet Archive. The values and references of the latter two quantities are noted in Table \ref{tab:planets_observations}.

\subsection{Sample preparation}
\label{sec:sampleprep}
We apply calibration steps to several Gaia DR3 data products. The effective temperature selected for use in this study is the `GSP-Spec' temperature, which is determined using the Matisse-Gauguin algorithm detailed in \cite{Creevey2022a} and \cite{recioblanco2016}, to derive atmospheric parameters from R$\sim$11,000 spectra observed with the RVS instrument on Gaia \citep{Cropper2018}. The lower and upper uncertainties are determined as the 16th and 84th percentile using Matisse-Gauguin and Monte Carlo realisation. As the distributions of the atmospheric parameters for all targets are overall symmetric and Gaussian, we simplify the description of them and symmetrise the uncertainty by using the mean of the two uncertainties. The reported uncertainties are small ($<30$~\si{\kelvin} in some cases) and do not take into account systematic errors. Following \cite{PerfVer22}, we add a systematic uncertainty of $90$~K in quadrature to the reported uncertainty, consistent with the expected uncertainties in effective temperature scales \citep{tayar22}.

A similar approach was taken with the errors for the metallicity, with a median scatter reported by \cite{Recio-Blanco2022a} of $0.13$~dex added in quadrature to the symmetrized error. We also corrected the metallicity values using the polynomial calibration over $\log g$ as described in  \cite{Recio-Blanco2022a}.

Finally, we corrected the measured parallax values for the known parallax zero-point offset in Gaia DR3 using the approach outlined in \cite{Lindegren2021} and \cite{LL:LL-124}.

\subsection{Determination of Stellar Parameters: \basta}
\label{sec:basta}

We make use of the BAyesian STellar Algorithm code \citep[\basta,][]{BASTA} to perform a homogeneous analysis of exoplanet host stars, alongside a custom computed set of stellar evolution models. Given a pre-computed grid of stellar models, \basta uses a Bayesian approach to compute the posterior distribution of a given stellar parameter using a set of observational constraints. It is fast and flexible and allows for a homogeneous characterisation even when the same quality or quantity of data is not available for all targets in a given sample.
\par
The Bayesian framework used is, briefly, as follows. If we treat $\Theta$ as the desired parameters, mass, radius and age, and $D$ as the input data, for example \teff, $\log g$, and \feh, then the likelihood of finding $\Theta$ given $D$ is
\begin{equation}
P(\Theta | D) = \frac{P(D | \Theta)P(\Theta)}{P(D)}
\end{equation}
The full likelihood is the product of the likelihood of groups of observables $D_i$:
\begin{equation}
P(D | \Theta) = \prod_i P (D_i | \Theta)
\end{equation}
where 
\begin{equation}
P(D_i | \Theta) = \frac{1}{\sqrt{2\pi |\Vec{C_i}|}} \exp(-\chi^2 / 2)
\end{equation}
Here $|\vec{C_i}|$ is the determinant of the covariance matrix, which is used to account for known correlations in parameters in the fitting procedure. The computed posterior is used to calculate the marginalized posterior for the stellar model parameter $\theta$, where $\Theta'$ is all model parameters excluding $\theta$, and $x_{\vec{\Theta}}$ is a bayesian weight used to account for the volume of paramter space a model characterised by $\Theta'$ encompasses.    \citep{BASTA}.
\begin{equation}
P(\theta | \vec{D}) = \int P(\theta , \vec{\Theta'} | \vec{D})x_{\vec{\Theta}} d\vec{\Theta'}
\end{equation} 
Quantities that are not directly observable like stellar age can then be inferred given the constraints from the observables.
\par
We determine the stellar properties using grid-based stellar modelling, comparing the observed parameters to the predicted theoretical quantities. Specifically, we use the astrometric coordinates (right ascension, declination), parallax ($\varpi$), metallicity (\meh), effective temperature (\teff), and Gaia magnitudes: G, Bp, and Rp, all from Gaia DR3, as inputs.
\par

 For stars with literature stellar masses $\stellarmass<1.15$~\si{\solarmass}, we build a grid with $\sim6000$ evolutionary tracks of stellar models. Our grids of stellar models were computed with the Garching Stellar Evolution Code \citep[\Garstec,][]{weiss2008}. The code utilises a combination of the equation of state by the \textsc{OPAL} group \citep{rogers1996,rogers2002}, and the Mihalas-Hummer-D\"{a}ppen equation of state \citep{mihalas1988,hummer1988,daeppen1988,mihalas1990}. We use the OPAL opacities \citep{rogers1992,iglesias1996} at high temperatures supplemented by the opacities of \citep{ferguson2005} at low temperatures. \Garstec uses the NACRE nuclear reaction rates \citep{angulo1999} except for $^{14}$N($p,\gamma$)$^{15}$O and $^{12}$C($\alpha,\gamma$)$^{16}$O for which the rates from \cite{formicola2004} and \cite{hammer2005} were used. The stellar models are computed using the \cite{asplund2009} solar mixture. 
\par
Convection in the models is parameterised using mixing-length theory \citep{bohm1958, kippenhahn2012}, where the mixing length parameter is allowed to vary in the range $1.6$--$2.0$. We sample the parameter space by utilising the Sobol quasi-random, low-discrepancy sequences to uniformly populate the parameter space \citep{sobol1,sobol2,sobol6,sobol4,sobol5,sobol3}. The stellar grid samples initial iron abundances from $-1.0$--$0.6$~dex, stellar masses from $0.7$--$1.2$~\si{\solarmass}, mixing-length parameter from $1.6$--$2.0$, and initial helium fraction from $0.24$--$0.32$. The stellar models include the effects of atomic diffusion. We compute the theoretical asteroseismic mode frequencies for the models using the Aarhus adiabatic oscillation package \citep[\adipls;][]{jcd2008}. For the computation of synthetic magnitudes, we use the bolometric corrections of \cite{hidalgo2018} and we use the 3-dimensional dust map from \cite{green2019} for computing the extinction along the line-of-sight for each target. \basta allows for prior probability distributions to be taken into account when computing the posterior distributions. We used the Salpeter initial mass function \citep{salpeter1955} as a prior to quantify the expected mass distribution of stars favouring low-mass stars as the most abundant. 
\par
\Garstec models are calculated along fixed iron abundances \feh with alpha-abundance element mixtures ranging from $\alphafe=-0.2$ to $\alphafe=0.6$. \basta adopts the following conversion to the bulk metallicity \meh~: 
\begin{equation}
\feh = \meh - \textup{corr}(\alphafe)
\label{equ:alpha}
\end{equation} 
where corr(\alphafe) is a correction factor determined as a function of the adopted solar mixture for the selected model, as in \cite{salaris1993}. We supply \meh input values from Gaia RVs spectra, which are then converted by \basta to \feh using Equation \ref{equ:alpha} for each \alphafe grid, therefore ensuring consistency in the atmospheric input parameters.
\par
For stars with literature $\stellarmass>1.15$~\si{\solarmass}, we build a similar grid. However, for stars with a mass limit of $1.1 < \stellarmass < 1.6$~\si{\solarmass} we do not include atomic diffusion but instead include the effects of core overshooting in the models. With this more appropriate input physics, we avoid underestimating masses and metallicities when fitting for more massive stars.
Some stars in our sample have been precisely characterized using asteroseismology \citep{Huber2013,SA2015}. In order to ensure consistency, we do not use literature analyses of these stars but instead include the measured global asteroseismic observables, the large frequency separation \deltanu and the frequency of maximum power \numax, in our inference of stellar properties for these targets. These additional constraints improve the precision on the stellar age for this subset of our sample. \citep{Soderblom2010}. %$\Delta \nu$ and $\nu_{\textup{max}}$ were used as inputs to \basta, in addition to the same spectroscopic and photometric inputs for all other stars, for the following targets: 
We adopted asteroseismic parameters for the following stars: Kepler-21, Kepler-25, Kepler-68, Kepler-93, Kepler-100, Kepler-103, Kepler-129 -- 3 of which are in the final rocky planet sample, and 4 of which are sub-Neptunes (see Section~\ref{sec:planet}). Input asteroseismic values were adopted from \cite{Huber2013}. For Kepler-10, previous stellar characterisation using asteroseismology has been carried out \citep{fogtmannschulz}, but no \numax has been reported. Hence, we did not include asteroseismic constraints of Kepler-10. 
\par
\begin{figure}%
\centering
\includegraphics[width=0.5\textwidth]{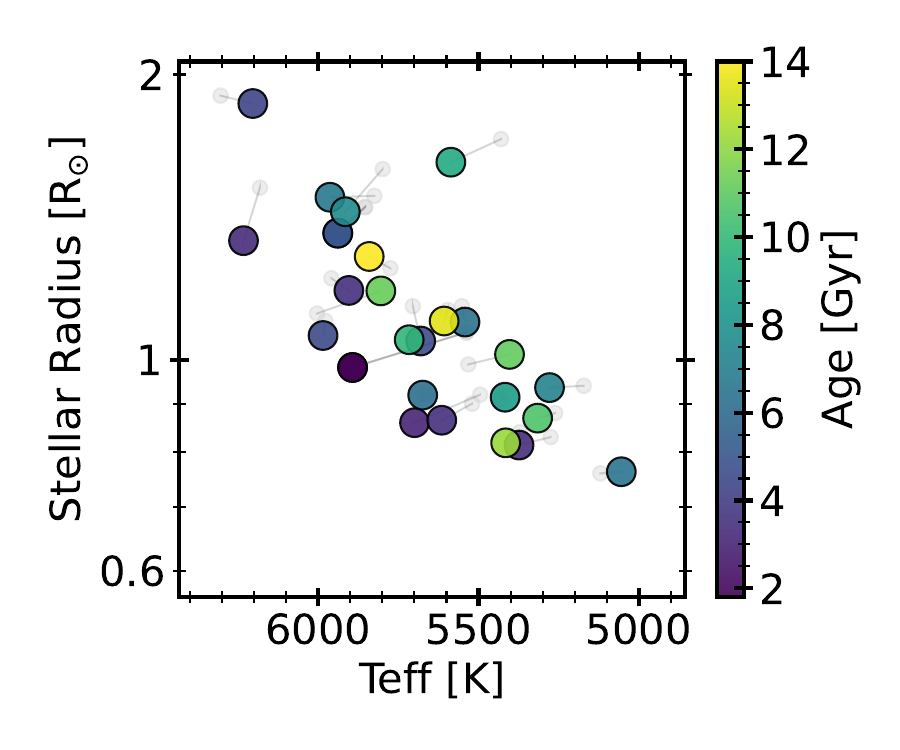}
\caption{{Homogeneous data from Gaia enables the precise relative measurement of ages for planet-hosting stars.} Stellar radius versus effective temperature for the 25 host stars in our sample that host rocky exoplanets. Points are colour-coded by the inferred stellar ages from \basta. Grey points connected via thin lines indicate previous, heterogeneously determined parameters for these host stars from the literature.% 
}
\label{fig:sample_teff_v_radius}
\end{figure}

Stellar parameters for our final sample are listed in Table~\ref{tab:hoststars_stellarproperties}. In Figure~\ref{fig:sample_teff_v_radius}, the stars are shown in a Hertzsprung-Russell diagram, demonstrating how comparing homogeneous constraints on effective temperature, radius, and metallicity from Gaia to stellar models enables precise relative measurements of ages for solar-type stars, which span from about $2$ to $14$~Gyr for our sample. These tests demonstrate the accuracy and precision of $\approx$2\,Gyr ($\approx$ 30\% for the median age in our sample) of ages in our sample. The precision is also consistent with our median estimated age uncertainties.

\par

\begin{table*}
\begin{tabular}{m{0.2cm}  m{3cm}  m{2cm} m{2cm} m{2cm} m{2cm}}
\hline
\hline
\\
%{} &           Planet name &  Stellar Mass [M$_{\odot}$] &  Stellar Radius [R$_{\odot}$] &  Teff [K] & [Fe/H] [dex] \\
{} &           Planet name &  Stellar Mass &  Stellar Radius &  \teff & \feh \\
{} &           &   [M$_{\odot}$] &   [R$_{\odot}$] &   [K] & [dex]
\\
\hline
\hline
1   &       Kepler-20 b &     $0.89^{+          0.07}_{-          0.06}$ &     $0.86^{+          0.03}_{-          0.03}$ & $ 5700^{+      46}_{-      95}$ &$-0.09^{+      0.17}_{-      0.14}$\\
2   &  EPIC 249893012 b &     $1.09^{+         0.09}_{-        0.07}$ &     $1.62^{+          0.06}_{-          0.07}$ &  $5586^{+      79}_{-     78}$ &  $0.20^{+      0.15}_{-      0.12}$ \\
3   &          K2-291 b &     $0.90^{+          0.07}_{-         0.05}$ &     $0.86^{+          0.03}_{-          0.03}$ &  $5615^{+      62}_{-      63}$ & $-0.03^{+      0.14}_{-      0.13}$ \\
4  &         WASP-47 e &     $1.06^{+          0.09}_{-         0.08}$ &     $1.10^{+          0.04}_{-          0.05}$ &  $5542^{+      73}_{-      73}$ &  $0.33^{+      0.12}_{-     0.08}$ \\
5  &  EPIC 220674823 b &     $0.93^{+          0.09}_{-         0.10}$ &     $1.01^{+          0.04}_{-         0.04}$ &  $5403^{+      86}_{-      87} $&  $0.16^{+      0.16}_{-      0.16}$ \\
6  &        HD 80653 b &     $1.17^{+          0.07}_{-          0.05}$ &     $1.18^{+          0.05}_{-          0.05}$ & $ 5904^{+      84}_{-      86}$ &  $0.33^{+      0.09}_{-      0.11}$ \\
7  &         CoRoT-7 b &     $0.86^{+         0.05}_{-          0.05} $&     $0.81^{+          0.03}_{-          0.03}$ & $ 5374^{+      45}_{-      72} $&  $0.05^{+      0.15}_{-      0.13}$ \\
8  &      Kepler-107 c &     $1.19^{+         0.08}_{-         0.06}$ &     $1.36^{+         0.09}_{-          0.08}$ &  $5939^{+     166}_{175}$ &  $0.34^{+      0.10}_{-      0.13}$ \\
9  &      Kepler-107 b &     $1.19^{+         0.08}_{-         0.06}$ &     $1.36^{+         0.09}_{-          0.08}$ &  $5939^{+     166}_{175}$ &  $0.34^{+      0.10}_{-      0.13}$\\
10  &          K2-131 b &     $0.78^{+          0.05}_{-          0.06}$ &     $0.76^{+          0.03}_{-          0.03}$ & $ 5055^{+      72}_{-      87}$ & $-0.03^{+      0.14}_{-      0.17}$\\
11  &         HD 3167 b &     $0.84^{+          0.07}_{-          0.07}$ &     $0.87^{+          0.03}_{-          0.04}$ & $ 5316^{+      81}_{-      68}$ & $-0.01^{+      0.14}_{-     0.16}$ \\
12  &          K2-111 b &     $0.80^{+          0.06}_{-          0.08}$ &     $1.2^{+          0.04}_{-          0.05}$& $ 5841^{+     101}_{-      70 }$& $-0.54^{+      0.19}_{-     0.21}$ \\
13  &      Kepler-100 b &     $1.08^{+          0.05}_{-          0.04}$ &     $1.49^{+          0.03}_{-          0.02}$ & $ 5964^{+      25}_{-     39}$ & $-0.00^{+      0.12}_{-      0.09}$ \\
14  &         TOI-561 b &     $0.77^{+          0.04}_{-          0.06}$ &     $0.82^{+          0.03}_{-          0.03}$ & $ 5415^{+      67}_{-      79}$ & $-0.27^{+      0.14}_{-      0.14}$ \\
15  &      Kepler-323 c &     $0.90^{+          0.13}_{-          0.17}$ &     $1.18^{+          0.05}_{-          0.06}$ & $ 5805^{+     10}_{-      86}$ & $-0.18^{+      0.39}_{-      0.34}$ \\
16  &           K2-38 b &     $1.04^{+          0.09}_{-          0.09}$ &     $1.05^{+          0.04}_{-          0.04} $&  $5680^{+      75}_{-      80} $&  $0.21^{+      0.16}_{-      0.14}$ \\
17  &        HD 20329 b &     $0.88^{+          0.08}_{-          0.10}$ &     $1.10^{+          0.04}_{-          0.04}$ &$  5607^{+      78}_{-      85}$ & $-0.09^{+      0.17}_{-      0.17}$ \\
18  &       Kepler-21 b &     $1.30^{+          0.10}_{-        0.07}$ &     $1.86^{+          0.05}_{-          0.05}$ &  $6204^{+      77}_{-      63} $& $0.13^{+     0.11}_{-      0.14}$ \\
19  &          55 Cnc e &     $0.94^{+          0.08}_{-         0.08}$ &     $0.94^{+          0.04}_{-          0.03}$ & $ 5279^{+      62}_{-      82}$ &  $0.26^{+      0.15}_{-      0.12}$ \\
20  &       Kepler-93 b &     $0.90^{+          0.04}_{-         0.05}$ &     $0.92^{+          0.01}_{-          0.02}$ & $ 5674^{+      57}_{-      58}$ & $-0.12^{+      0.11}_{-      0.14}$ \\
21  &       Kepler-10 b &     $0.91^{+          0.09}_{-        0.11 }$&     $1.05^{+         0.04}_{-          0.04}$ &  $5716 ^{+     88}_{-      84}$ & $-0.09^{+      0.18}_{-      0.18}$ \\
22  &          KOI-94 b &     $1.21^{+          0.09}_{-          0.07}$ &     $1.34^{+          0.09}_{-          0.08}$ & $ 6232^{      161}_{-      203}$ & $0.2^{+      0.14}_{-      0.17}$ \\
23  &      Kepler-406 b &     $1.02^{+          0.07}_{-          0.06}$ &     $0.98^{+         0.04}_{-          0.03}$ & $ 5893^{+      57}_{-     102}$ &  $0.08^{+      0.16}_{-      0.14}$ \\
24  &        TOI-1444 b &     $0.89^{+           0.08}_{-          0.08}$ &     $0.91^{+          0.03}_{-          0.03}$ &$  5417^{+      74}_{-      61}$ &  $0.06^{+      0.17}_{-      0.14}$ \\
25  &       HD 213885 b &     $1.00^{+          0.09}_{-          0.09}$ &     $1.06^{+          0.04}_{-          0.04}$ &$  5985^{+      82}_{-      75 }$& $-0.07^{+      0.18}_{-      0.16 }$\\
26 &       HD 137496 b &     $1.05^{+          0.10}_         {-0.09}$ &    $1.43^{+          0.05}_{-         0.05}$ &  $5915^{+      85}_{-      81 }$& $-0.04^{+      0.18}_{-      0.16}$ \\
\\
\bottomrule
\end{tabular}
\caption{\basta output stellar parameters for host stars of the 26 rocky planets in our sample -- the sample for which the composition-age trend is observed.}
\label{tab:hoststars_stellarproperties}
\end{table*}

We have tested our stellar mass, radius and age estimates in a various ways. We summarise our validation tests in the Appendices. In Appendix A1, we validate stellar mass and radius against literature values, and by using alternative spectroscopic inputs. Our stellar ages are validated in Appendix A2, where we compare results from our method to those of asteroseismology, gyrochronology, kinematics, and detailed abundances. We also apply this validation to a gold-standard sample of Kepler stars studied with asteroseismology.

\section{Planet sample and parameters}
\label{sec:planet}

%\subsection{Revised Planet Parameters \& Final Sample Selection}
%\label{sec:finalsample}

Once the new, homogeneous set of stellar parameters were obtained, and validated, we recalculated a new value of radius and mass for each planet. Planet radius is measured indirectly, by using the depth of the transit and the stellar radius as follows:

\begin{equation}
 \Delta F \approx \delta \approx \left(\frac{R_{p}}{R_{\star}}\right)^2,
\label{eq:relativeflux}
\end{equation}
where $R_p$ and $R_{\star}$ are planet and star radius respectively, and  $\Delta F$ is the observed reduction in relative flux, which is equal to transit depth $\delta$.
We use the reported values of $\frac{R_P}{R_{\star}}$ , which have taken into account limb darkening effects, and our updated value of $R_{\star}$ in order to calculate a new $R_p$.

\par

Planet mass is also recalculated using the new homogeneous stellar mass. We follow the same method as described in \cite{RADVEL}. In short, we substitute our new value for $M_{\star}$ into 

\begin{equation}
    M_p\sin i = \frac{K}{K_0} \sqrt{1-e^2} \frac{M_{\star}}{M_{\odot}} \frac{2}{3} P^{\frac{1}{3}}, 
\label{eq:mpsini}
\end{equation}
where $M_p$ denotes planet mass, $K$ the radial velocity semi-amplitude, $i$ the inclination, $e$ the eccentricity of the orbit, $M_{\star}$ the stellar mass, $M_{\odot}$ the Solar mass and $P$ the orbital period. These values, except for $M_{\star}$, are extracted from the NASA exoplanet archive. 
\par
In Table~\ref{tab:planets_observations}, we list the adopted transit and RV observables used to update our planet parameters, in addition to the sources for these values.  In 15/26 ($\sim$57$\%$) of cases, $K$ and $\frac{R_P}{R_{\star}}$ are determined in the same paper.
\par
In the case of Corot-7 b, several literature studies have arrived at significantly differing values for the RV amplitude. We opt to use the values reported by \cite{john2022}, as this is the most recent study and it applies state-of-the-art stellar activity mitigation in its RV analyses. In all plots, we use values from this paper as the literature values. 
\par
At this stage, we use the inferred planet radius measurements to determine which planets are super-Earths located below the radius valley, and limit all further analysis to these planets only. To do so, we use the location of the radius valley as determined by \cite{Ho+Vanylen2022}.  
Specifically, the location of the radius valley depends on orbital period and stellar mass, as $\log (R_p/R_\oplus) = -0.09 \log (P/\mathrm{days}) + 0.21 \log (M_{\star}/M_{\odot}) + 0.35$ \citep[Eq.~11,][]{Ho+Vanylen2022}. For each planet in our sample, we calculate the location of the valley at its orbital period and stellar mass, and keep those planets whose radius is smaller than the valley radius.
Finally, we exclude three planets which have only reported mass upper limits, and thus no or uninformative mass errors. We exclude Kepler-131c, as its mass is non-physically large and poorly constrained as previous studies have noted \citep{Weiss+Marcy2014}.Kepler-406c is also excluded, due to having an unphysical density, calculated based on a tentative RV curve in \cite{Weiss+Marcy2014}. The K2-32 system is a close-in binary \citep{Mugrauer2019} whose Gaia parameters, in particular \meh are clearly very discrepant with the literature values, so this star is removed from the analysis. 
\par

\begin{table*}
\begin{tabular}{  m{0.2cm}  l l l r l}

\hline
\hline
%{} &           Name &  K [m/s] &                                   K Reference &  $R_p / R_{\star}$ &                                  Reference \\
{} &           Planet name &  K &                                   Ref &  $R_p / R_{\star}$ &                                  Ref \\
{} &            &  [m/s] &                                    &   &   \\                       
\hline
\hline
1 &       Kepler-20 b &      $4.23^{+0.54}_{-0.54}$ &  A &    $0.01925^{+0.00018}_{-0.00018}$ &  C \\
2  &  EPIC 249893012 b &      $3.55^{+0.43}_{-0.43}$ &  B &    $0.01040^{+0.00040}_{-0.00040}$ &  B \\
3  &          K2-291 b &      $3.33^{+0.59}_{-0.59}$ &  D &    $0.01614^{+0.00062}_{-0.00033}$ &  D \\
4  &         WASP-47 e &      $4.55^{+0.37}_{-0.37}$  &  E &    $0.01458^{+0.00013}_{     -0.00013}$ &  E\\
5  &  EPIC 220674823 b &      $6.50^{+0.52}_{         -0.52}$ &  A &    $0.01601^{+0.00031}_{       -0.00029}$ &  F \\
6 &        HD 80653 b &      $3.62^{+0.21}_{-0.21}$ &  A &    $0.01210^{+ 0.00040}_{-0.00040}$ &  G \\
7 &         CoRoT-7 b &      $3.10^{+0.70}_{-0.70}$ & H &    $0.01870^{+0.00030}_{-0.00030}$ &  H \\
8 &      Kepler-107 c &      $3.29^{+0.66}_{-0.66}$ & A&    $0.01012^{+0.00013}_{-0.00013}$ & A\\
9 &      Kepler-107 b &      $1.44^{+0.68}_{-0.68}$  &  A &    $0.00973^{+0.00013}_{-0.00013}$ &  A\\
10  &          K2-131 b &      $8.00^{+1.30}_{-1.30}$ &  A &    $0.01720^{+0.00041}_{-0.00027}$ &  I\\
11  &         HD 3167 b &      $3.56^{+0.15}_{-0.15}$ & A &    $0.01725^{+0.00023}_{       -0.00023}$ &  J \\
12  &          K2-111 b &      $2.30^{+0.30}_{-0.30}$ & A &    $0.01255^{+0.00050}_{      -0.00048}$ &  K \\
13 &      Kepler-100 b &      $1.90^{+0.80}_{-0.80}$ & L &    0.00809 & L\\
14 &         TOI-561 b &      $2.18^{+0.20}_{-0.20}$ &  M &   $0.01550^{+0.001}_{-0.001}$ &  M\\
15 &      Kepler-323 c &      $2.80^{+1.40}_{-1.40}$&  A &    $0.01337^{+0.00023}_{-0.00023}$ &  N\\
16 &           K2-38 b &      $3.02^{+0.43}_{-0.43}$ &  A&    $0.01330^{+0.00071}_{-0.00076}$ &  O\\
17 &        HD 20329 b &      $5.07^{+0.41}_{-0.42}$ & P &    $0.01390^{+0.00050}_{-0.00050}$ & P\\
18 &       Kepler-21 b &      $2.70^{+0.46}_{-0.46}$ &  A &    $0.00788^{+0.00005}_{-0.00005}$ &  Q \\
19 &          55 Cnc e &     $ 6.02^{+0.24}_{-0.23}$&  S &    $0.01820^{+0.00020}_{-0.00020}$& S\\
20 &       Kepler-93 b &      $1.89^{+0.21}_{-0.21}$ &  A&    $0.01475^{+0.00006}_{-0.00006}$&  T\\
21 &       Kepler-10 b &      $2.34^{+0.21}_{-0.21}$ &  A &    $0.01268^{+0.00004}_{-0.00004}$  &  U\\
22 &          KOI-94 b &      $3.30^{+1.40}_{-1.40 }$ & R &    $0.01031^{+0.00014}_{-0.00014}$ & R\\
23 &      Kepler-406 b &      $2.89^{+0.60}_{-0.60}$&  L &    0.01229 & L  \\
24 &        TOI-1444 b &      $3.30^{+0.58}_{-0.59}$ &  V &    $0.01410^{+0.00060}_{-0.00058}$  &  V\\
25 &       HD 213885 b &      $5.30^{+0.39}_{-0.39}$ & W &    $0.01453^{+0.00041}_{-0.00042}$ & W\\
26 &       HD 137496 b &      $2.14^{+0.29}_{-0.29}$ & X &    $0.0076^{+0.00030}_{-0.00029}$ & X\\
\\
\bottomrule
\end{tabular}

\caption{RV semi-amplitude ($K$), used to re-calculate planet masses, and transit-derived planet-to-star ratio (${R_p}/{R_\star}$), used to re-calculate planet radius, with references for each as follows: A - \citet{Bonomo23}, B - \citet{hidalgo2020},
C - \citet{Burke2014},
D - \citet{Kosiarek2019},
E - \citet{Bryant2022},
F - \citet{Guenther2017},
G - \citet{Frugastili2020},
H - \citet{john2022},
I - \citet{Adams2021},
J - \citet{Vanderburg2016},
K -  \citet{Fridlund2017},
L - \citet{Marcy2014},
M - \citet{brinkman23},
N - \citet{Burke2014},
O - \citet{Toledopadron2020},
P - \citet{Murgas22},
Q - \citet{Lopez-Morales2016},
R - \citet{Weiss2013},
S - \citet{Bourrier2018},
T - \citet{Dressing2015},
U - \citet{Dai2019} ,
V - \citet{Dai2021} ,
W - \citet{Espinoza2020}
X - \citet{Azvedo2022}.
}
\label{tab:planets_observations}
\end{table*}

\section{A link Between Age and Composition}
\label{sec:agecomposition}
\par
\begin{figure*}%
\centering
\includegraphics[width=0.9\textwidth]{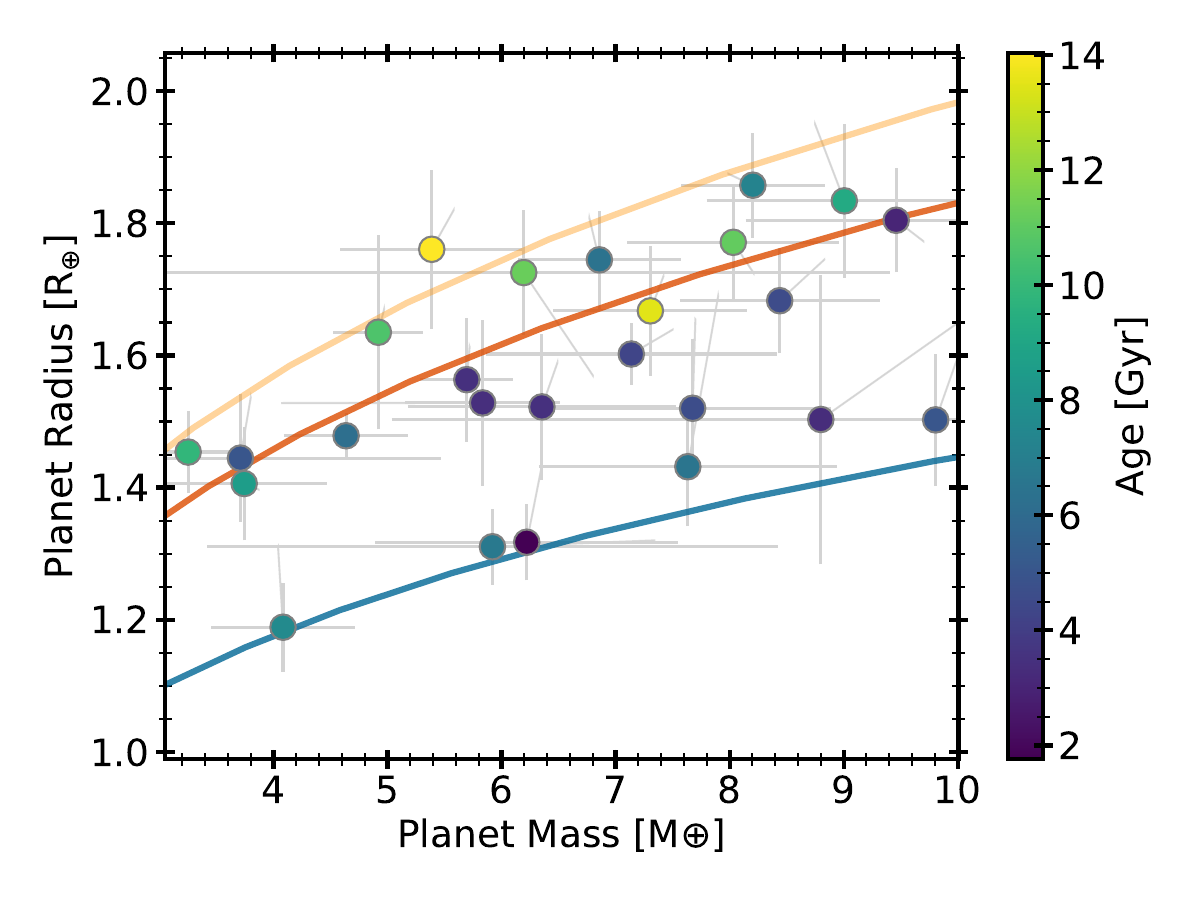}
\caption{{Rocky planet density and composition depends on the age of the host star.} The mass and radius of small planets, colour-coded by age, alongside composition tracks from \citet{Zeng2016}. The upper, light orange line corresponds to a planet with composition of 100$\%$ MgSiO$_3$, i.e. 100$\%$ rocky mantle. The middle, dark orange line corresponds to a planet with 67.5$\%$ MgSiO$_3$, and 32.5$\%$ Fe, providing an analogue for Earth-like composition. The lower blue line represents planets which would be composed of 100$\%$ Fe, essentially pure iron cores.  Grey lines indicate movement from previously reported, heterogeneous values from the NASA Exoplanet Archive.}
\label{fig:sample_rockyplanetdensity_dependson_age}
\end{figure*}

In Table ~\ref{tab:planets_properties}, we list our adopted planet parameters. The planets in our sample are shown in a mass-radius diagram in Figure~\ref{fig:sample_rockyplanetdensity_dependson_age}. Models for rocky planets with varying compositions \citep[taken from][]{Zeng2016} are also shown.  These tracks are created from semi-empirical relations derived under isothermal formation scenarios for rocky exoplanets with gaseous envelopes of negligible mass fraction. They represent rocky planets at different percentages of mantle materials such as MgSiO$_3$ and MgSiO$_4$ with respect to iron, which is assumed to be the main constituent of a core \citep{Seager2007}. In Fig.~\ref{fig:sample_rockyplanetdensity_dependson_age}, we furthermore colour-code each planet by its age, where we have assumed that planet ages are identical to the age of their host stars, since studies of young stellar objects indicate that the protoplanetary disk phase lasts only a few million years following the collapse of their parent molecular cloud \citep{Haisch2001,Hartmann1998}. Visually, this seems to reveal a correlation between planet composition and stellar age. In the age range of the sample, younger stars appear more likely to host denser, more iron-rich rocky planets. Planets above the `Earth like' composition track in  Fig.~\ref{fig:sample_rockyplanetdensity_dependson_age} seem to orbit about older stars than those below it. We observe a hitherto unknown trend that rocky planets which are denser and have a higher inferred Fe content appear to orbit stars that are younger, suggesting a link between the composition of rocky planets and their age. We now investigate this trend in more detail.

\subsection{Planet density and inferred iron fraction}
\label{sec:fefraction}

\begin{figure}
\centering
\includegraphics[width=0.45\textwidth]{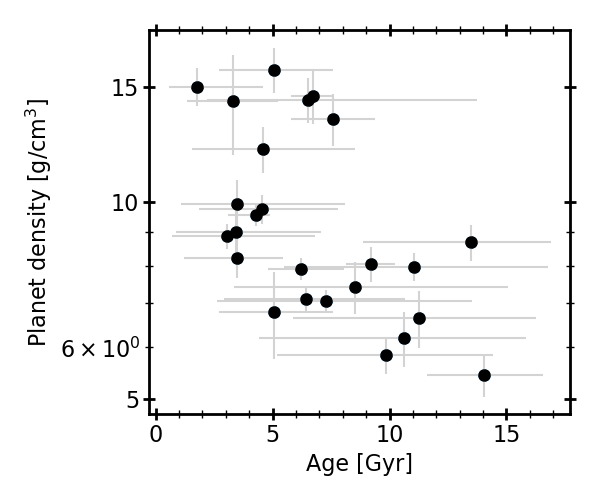}
\caption{Planet density, in log scaling on the y-axis, against stellar age on the x-axis, for planets and stars in our rocky planet sample. A clear trend can be seen, with denser planets orbiting around younger stars.}
\label{fig:density-age}
\end{figure}

In Fig.~\ref{fig:sample_rockyplanetdensity_dependson_age}, we showed a mass-radius diagram with planets colour-coded by age. From this diagram, a visual correlation between planet composition and age appears to be apparent. An alternative way of showing this effect is to plot the stellar density as a function of age, which is what we do in Fig.~\ref{fig:density-age}, which similarly reveals a visual correlation between planet density and age.

%\par

\begin{figure}
\centering
\includegraphics[width=0.5\textwidth]{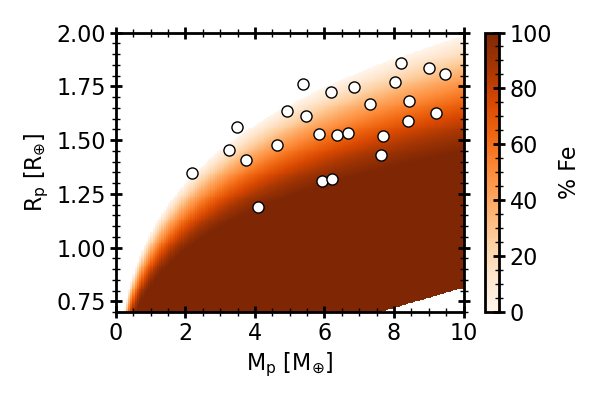}
\caption{Interpolated grid of iron mass fraction values for planets at a given mass and radius, measured in Earth-scaled values. The colour bar represents the percent mass fraction of iron, as interpolated from Table~2 of \citet{Zeng2016}. White overlayed points highlight our sample of rocky planets.}
\label{fig:significance_interpolatedgrid_ironmassfraction}
\end{figure}

To quantify this trend, we first note that planet density is an incomplete metric for measuring planet composition. This is because planets with the same density do not necessarily have the same composition, as planet composition tracks do not follow equal-density tracks. 
%
%\subsection{Calculation of Planet Iron Fractions}
%
To address this, we can infer the planet composition using a grid of planet interior models \citep[][Table 2 therein]{Zeng2019}, which gives values for radii as a function of mass, at incremental values of iron percent. This provides a sparse grid of points for which values of mass, radius, and iron mass fraction are available. We take this grid and linearly interpolate in two dimensions, namely in mass-radius space, using \verb+`scipy.interpolate.LinearNDInterpolator'+. The newly interpolated grid that we create is shown in Fig.~\ref{fig:significance_interpolatedgrid_ironmassfraction}. In order to assign a value of iron fraction to each planet in our sample, we draw 100 random samples from a Gaussian distribution with $\mu$, $\sigma$ treated as the measurement and uncertainty respectively, for planet mass and radius. For each random tuple, a value of \ironmassfrac is calculated from the interpolation, and these 100 values are used to create a posterior distribution function for the \ironmassfrac. We take the median of \ironmassfrac as the final value, and lower and upper uncertainties as the 16th and 84th percentiles, respectively. Planets with best-fit masses and radii that fall above the `pure-rock' line in are assigned a value of $\ironmassfrac = 0$. While this may not be a realistic physical description of the planet compositions, we note that all such planets are consistent with pure-rock models within their uncertainties, and that some of this discrepancy may be due to systematic errors in the models.

\par

\subsection{Significance of planet composition-age trend}
\begin{figure*}%
\centering
\includegraphics[width=0.9\textwidth]{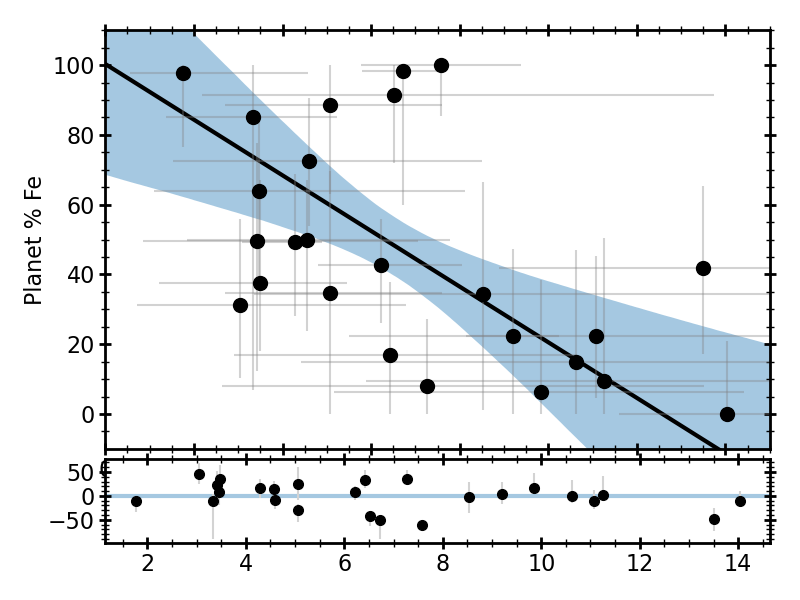}
\caption{\textbf{Younger, denser rocky planets have higher iron content}. Calculated iron mass percentages of planets, based on their measured masses and radii and models interpolated from a grid by \citet{Zeng2019}, shown as a function of age. A Bayesian linear regression yields a
slope and intercept of $\alpha =  -8.0^{+4.6}_{-5.0}$,
$\beta =  100^{+34}_{-32}$, and a median correlation coefficient of $-0.62^{+0.23}_{-0.17}$.
The solid black line shows the best-fit model, and the blue shaded area is the 1- $\sigma$ confidence interval. The residuals of the model are shown in the bottom panel.}
\label{fig:sample_youngerplanets_higheriron}
\end{figure*}

In Fig.~\ref{fig:sample_youngerplanets_higheriron}, we show the inferred \ironmassfrac of the planets (see Section~\ref{sec:fefraction}) as a function of stellar age ($\tau$). To quantify this trend, we first calculate correlation coefficients. We find Pearson, Spearman, and Linear Least Squares regression coefficients that are $-0.62$, $-0.63$, $-0.62$, suggesting a negative correlation where the planet iron percentage decreases with stellar age. 

\begin{figure*}
\centering
\includegraphics[width=0.9\textwidth]{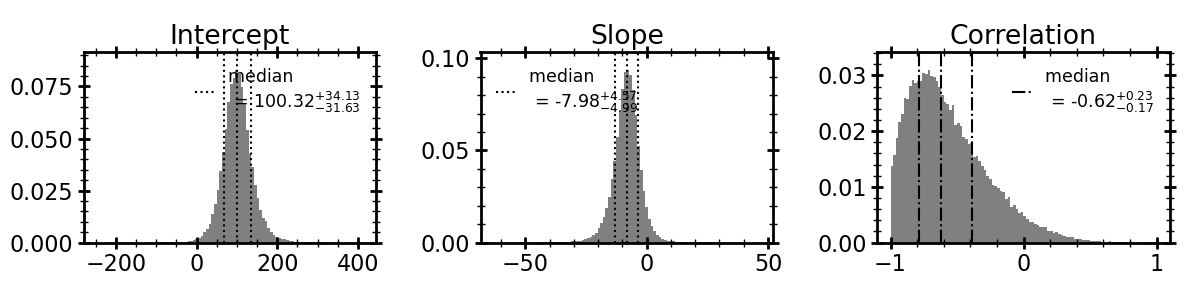}
\caption{The two leftmost panels show posteriors of the intercept and slope -- defined as $\alpha$ and $\beta$ in our model, of the fit between age and planet iron mass fraction \ironmassfrac using \linmix \citep{Kelly2007}. In the rightmost panel, a PDF of the Pearson correlation coefficient is plotted, showing a clear negative linear correlation between age and \ironmassfrac. These values were calculated using a Bayesian Linear Regression fit.}
\label{fig:significance_alpha_beta_corr}
\end{figure*}

In order to further quantify the observed trend, we fitted the observed age-planet iron trend using a linear model $\tau = \alpha \ironmassfrac + \beta$. We did so using \linmix \citep{Kelly2007}, which performs a Bayesian linear regression, taking into account the errors on both $x$ and $y$ (i.e., age and iron content). We find $\alpha =  -8.0^{+4.6}_{-5.0}$,
$\beta =  100^{+34}_{-32}$, and a median correlation coefficient of $-0.62^{+0.23}_{-0.17}$ .Figure~\ref{fig:significance_alpha_beta_corr} shows the posterior distributions for $\alpha$, $\beta$ and the correlation coefficient. The correlation is highly significant.
%and the modelled slope is significant at a level greater than 2-$\sigma$. 
%

\begin{figure}
     \centering
     \begin{subfigure}{0.45\textwidth}
     \centering
         \includegraphics[width=\textwidth]{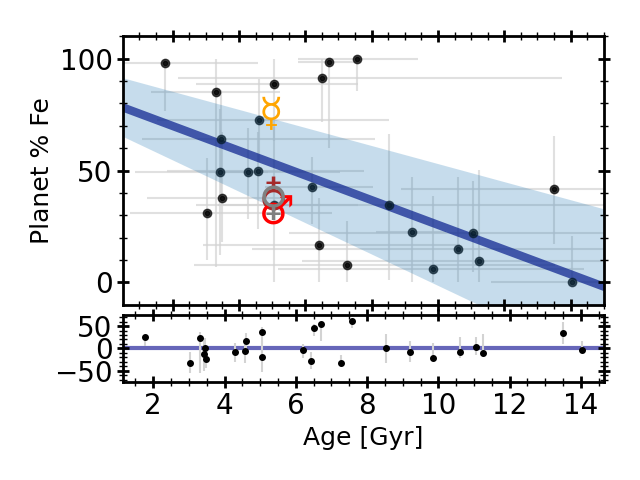}
     \end{subfigure}
    \hfill
     \begin{subfigure}{0.45\textwidth}
     \centering
         \includegraphics[width=\textwidth]{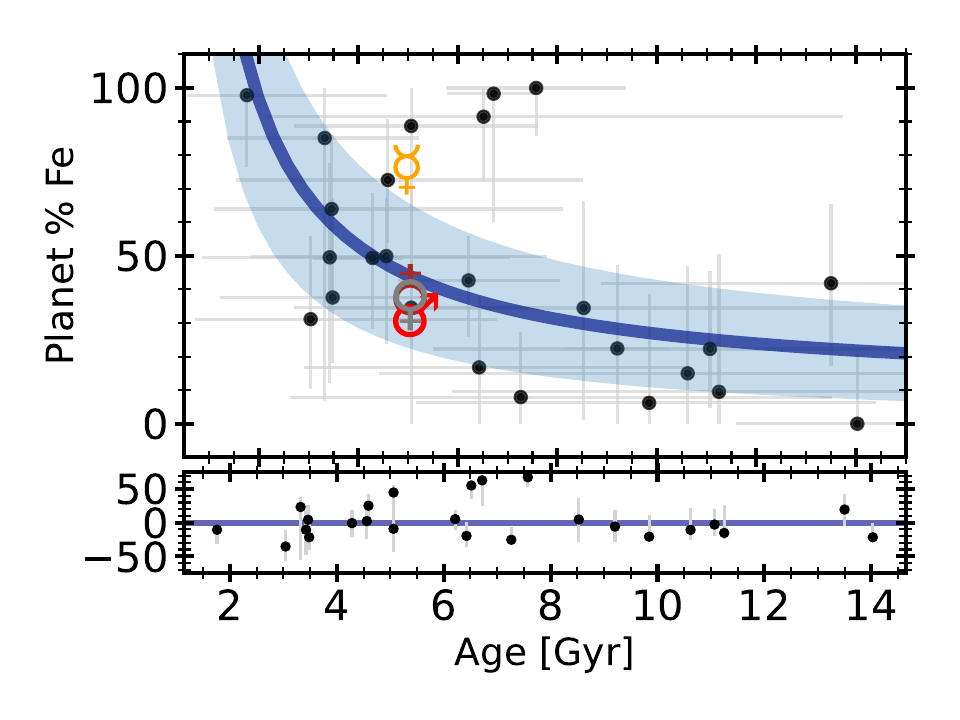}
     \end{subfigure}

\caption{Calculated iron mass percentages of planets, based on their measured masses and radii and models interpolated from a grid by \citet{Zeng2019}, shown as a function of age. Upper: An orthogonal distance regression using a linear model of the form $\alpha x + \beta$ is plotted in dark blue, atop the shaded $1-\sigma$ confidence region in light blue. Lower: An ODR using an inverse linear model of the form $ ( \gamma / x ) + \eta$ is plotted in dark blue, atop the shaded $1-\sigma$ confidence region in light blue. The Solar system planets are plotted, with Earth, Venus and Mars having similar \ironmassfrac, between 25 and 40. Mercury has ~$70\%$ Fe. Solar system values used are from \citep{McDOnough2021}, and the residuals of both models are shown in the respective bottom panels.}
    \label{fig:significance_ironmasspercentages_from_interpolatedgrid}
\end{figure}

To explore alternative methods, we fit our data using an Orthogonal Distance Regression Method \citep{ODR}, using \verb+scipy.odr+. This method takes into account errors on age and $\%$ Fe, and can fit to non-linear functional forms. We heuristically choose two models, shown respectively in Fig.~\ref{fig:significance_ironmasspercentages_from_interpolatedgrid}. First, we fit a linear model of the form $\alpha x + \beta$, which is plotted in dark blue, atop the shaded $1-\sigma$ confidence region in light blue. With this method, $\alpha = -5.52 \pm 1.49$ and $ \beta = 81.0 \pm 12.2$, which agree, within errors, to our values from the Bayesian linear regression discussed above, and also suggest a non-zero slope at a significance of more than $2 \sigma$. We repeat this method using an inverse linear model of the form $ ( \gamma / x ) + \eta $, giving values of $\gamma = 174 \pm 55$ and $\eta = 9.35 \pm 10.5$. Both methods show correlation between the two parameters, confirming the results derived using \linmix in Fig.~\ref{fig:sample_youngerplanets_higheriron}.

\subsection{Investigating alternative correlations}
At a given stellar radius, temperature  and metallicity, a higher-mass star will have a lower age than a lower-mass star. Therefore we investigate whether the observed trend with stellar age could be caused by a physical effect that is related to stellar mass.
\begin{figure}
\centering
\includegraphics[width=0.5\textwidth]{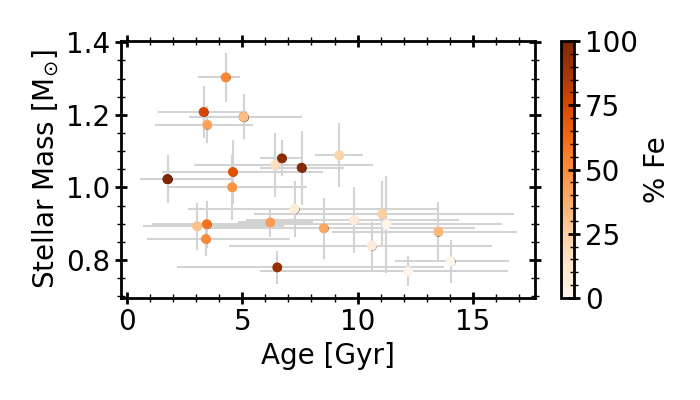}
\caption{Stellar mass versus age for stars in our rocky planet sample. The spread of ages is larger for lower-mass stars, which is expected as such stars live longer. Data points are colour-coded by \ironmassfrac, demonstrating a visual gradient which is stronger with age (from left to right) than with mass (from bottom to top).}
\label{fig:investigating_stellarmass_v_age}
\end{figure}

Fig.~\ref{fig:investigating_stellarmass_v_age} shows stellar age versus mass for the stars in our sample, colour-coded by their inferred planet \ironmassfrac. This plot shows that higher-mass stars span a narrower range of ages and are mostly younger relative to their lower-mass counterparts, as expected. Visually, a trend of \ironmassfrac with age can be seen more clearly than a trend with stellar mass. An age trend can be seen most clearly at $M_s < 1 M_{\odot}$, while the data is consistent with such a trend at higher masses too.
\begin{figure*}
\centering
\includegraphics[width=\textwidth]{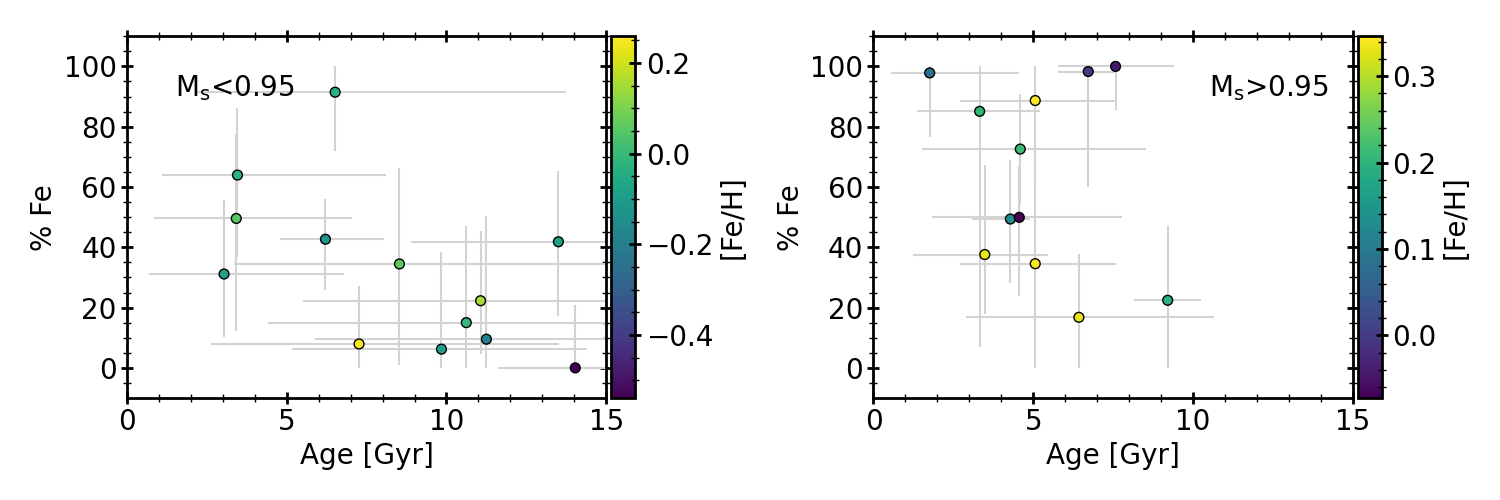}
\caption{\ironmassfrac plotted as a function of stellar age, for two subsets of our sample: stars with $\stellarmass<0.95$~\si{\solarmass} (left), and $\stellarmass>0.95$~\si{\solarmass} (right). The age trend can be seen clearly in the left panel. The trend is weaker in the right-hand panel, which contains fewer stars of older ages.  The points are color-coded by stellar metallicity. While on average less massive stars are older and slightly metal poorer than more massive stars, we do not observe a clear trend between planet \ironmassfrac and stellar metallicity.}
\label{fig:investigating_ironmassfraction_v_mass}
\end{figure*}
\par To further explore the effect of stellar mass, in Fig.~\ref{fig:investigating_ironmassfraction_v_mass} we show the planet \ironmassfrac as a function of age, in two bins of stellar mass, split at $M_s=0.95$~\si{\solarmass} to contain a roughly equal part of our sample. The trend between stellar age and \ironmassfrac can be clearly seen within the lower mass bin, and less clearly in the higher mass bin, which has a narrower age dispersion. The points are color-coded by metallicity, again showing a lack of correlation between stellar metallicity and \ironmassfrac. We note that the dearth of high density (high \ironmassfrac) planets around old, high-mass stars may be caused by detection bias, since some of these stars will have appreciably evolved off the main-sequence. 
%It is possible that planets below 7-8 Gyr are more dense around higher mass stars than around low mass stars, but we caution this may be the result of small-number statistics. In that age range, only 5 planets in our sample are < 0.95 M$_\odot$, whereas there are 10 in the > 0.95 M$_\odot$ sample. Younger, higher mass stars do not host any planets with $\ironmassfrac < 20$.}

\begin{figure}
\centering
\includegraphics[width=0.5\textwidth]{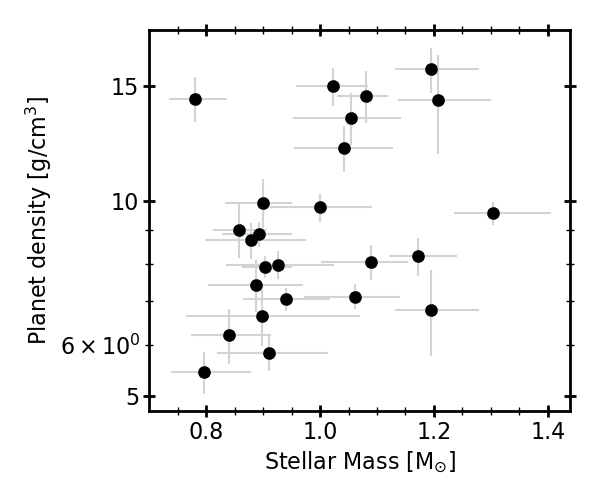}
\caption{Stellar mass as a function of planet density. A weak correlation is seen, as expected due to the inverse correlation between stellar mass and age. This is weaker than the trend observed between age and density.}
\label{fig:stellar_mass_v_density}
\end{figure}

We finally plot planet density as a function of stellar mass in Fig.~\ref{fig:stellar_mass_v_density}. A weak correlation is observable, with a Pearson correlation coefficient $r=0.35$, with P value of $0.08$), as expected due to the inverse correlation between stellar mass and age. This is weaker than the trend observed between age and density. This may be expected as on average, more massive stars are younger (see Figure~\ref{fig:investigating_stellarmass_v_age}). However, as not all old stars are less massive, this may explain why the observed correlation with stellar density appears weaker than that observed with stellar age. The latter is more strongly correlated with alpha-abundances \alphafe, which change with Galactic evolution, and which we expect to have a more measurable impact on planet composition \citep[i.e.][]{santos2017}.

\begin{figure}
\centering
\includegraphics[width=0.5\textwidth]{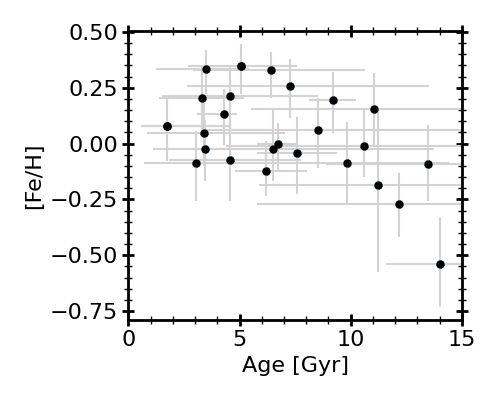}
\caption{Stellar metallicity \feh as a function of stellar age. No significant correlation between age and metallicity is observed, except for the highest ages, in agreement with literature.}
\label{fig:investigating_metallicity_v_age}
\end{figure}

We also plot stellar age as a function of \feh. Studies of Galactic stellar populations have shown that the age-metallicity relation is mostly flat until about 10~Gyr \citep[e.g.,][]{Haisch2001,snaith2015}. This is also what we observe, as we see no significant correlation between metallicity and age except at the highest ages, above approximately 10~Gyr, as shown in Figure~\ref{fig:investigating_metallicity_v_age}. This explains why we do not see a strong correlation between planet composition and stellar \feh. The observed planet density-age trend seems to instead show the greatest rate of change at around 6-8~Gyr.

\par
This is the first time that a link between the composition of rocky planets and their age is observed. Despite large uncertainties in age, this is the first homogeneous set of ages for a sample of such planets, using the same stellar models, input parameters, and astrometry. Therefore, we interpret age as an indirect link to terrestrial planet density as a result of its direct link to the chemical composition of the host star.

\subsection{Potential sample biases}
We consider whether the observed planet density versus age correlation could be affected by observational biases. 
In Fig.~\ref{fig:effects_everthing_v_age}, we plot a range of parameters as a function of stellar age, such as planet radius, planet mass, orbital period, stellar mass, stellar radius, and stellar magnitude. Although the stellar parameters are calculated homogeneously in this work, a homogeneous re-calculation of planet RV semi-amplitudes and transit depths is beyond its scope. Such heterogeneity may create biases in planet parameter space. Many different facilities and missions are used for these measurements, thus we expect that such biases, should they exist, to be stochastic, and unlikely to systematically affect the observed trend.

\subsubsection{Sample selection: transit detection.}
Firstly, we consider whether the sample selection could be biased. As most planets in our sample were first discovered by transit, we first consider whether transit detectability could influence our results. From the top left panel of Fig.~\ref{fig:effects_everthing_v_age}, it can be observed that the planet size distribution does not appear to depend on age, suggesting planet transits can be detected across the entire age range within our sample. While there may be reasons to expect that small planets are more often missed around young stars because such stars are more active, such an effect is expected to be unimportant at main sequence ($\geq$ 1~Gyr) ages. %Furthermore, if small planets, which are typically denser, are more likely to be missed around young starsthat would only enhance the true correlation between stellar age and planet density we observed, strengthening our conclusions.

We conclude that it is unlikely that transit detection has a significant impact on our conclusions.

\subsubsection{Sample selection: RV observations}
To be included in our sample, transiting planets require mass measurements, obtained using RV observations. Similar to with transit observations, it is a priori plausible that lower-mass planets are missed around the youngest stars, as these stars may be more active. Such an effect is not observed in the top right panel of Fig.~~\ref{fig:effects_everthing_v_age}. While stellar rotation or activity are a function of age, they primarily affect achievable RV precision for very young ages ($<500$~Myr) which are not present in our sample. The effect on RV precision across main sequence ages is small. For example, \cite{Isaacson2024} suggest that stars with values for the chromospheric activity indicator $R'_{HK} < 4.8$ are considered inactive. As shown in \cite{Huber22}, $R'_{HK}$ is generally below such values for stars above about 2~Gyr. With the exception of Kepler-406 (which hosts a low-mass planet), all stars in our sample are older than this. Nevertheless, there may be a decreasing sensitivity to lower density planets towards younger ages (e.g., the lower-left region on Figure~\ref{fig:sample_youngerplanets_higheriron}), as higher stellar activity levels affect the mass determination more than the transit detection in such systems. This may affect the shape or strength of the observed trend.

%but we expect it not to significantly impact our main conclusions given the age range of our sample.}

\subsubsection{Other effects}
\begin{figure*}
\centering
\includegraphics[width=0.6\textwidth]{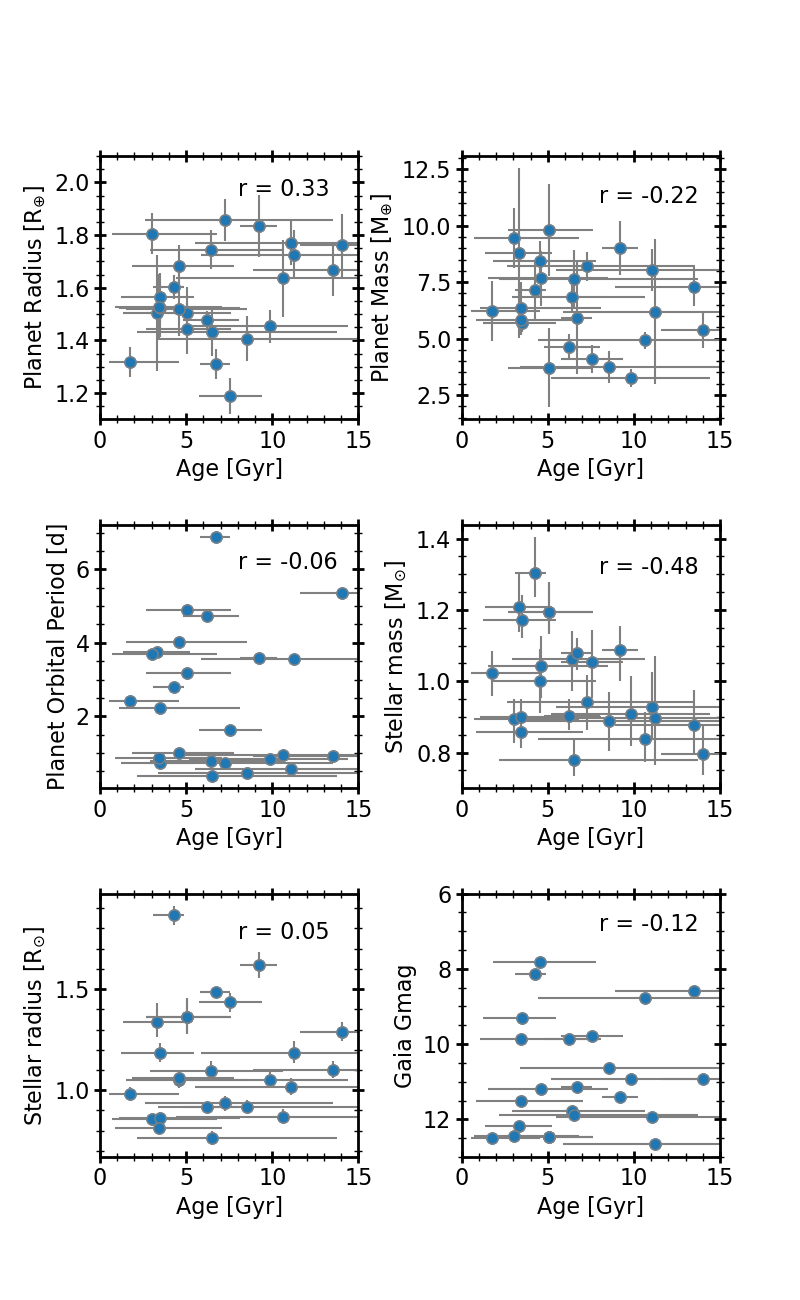}
\caption{Planetary and stellar parameters as a function of age for the 25 stars and 26 planets in this sample. Pearson correlation coefficients $r$ are quoted in each panel to indicate the calculated monotonic relationship between the variables.}
\label{fig:effects_everthing_v_age}
\end{figure*}

We consider whether other effects could influence our results. In the bottom right panel of Fig.~\ref{fig:effects_everthing_v_age}, we show the observed sample as a function of age and stellar magnitude. No clear correlation is observed, suggesting again that our sample of planets with both transits and RV observations is primarily a magnitude-limited sample that is not selected as a function of age (which would have largely been unknown to observers at the time of detection or follow-up). 
A correlation between stellar mass and age can be observed as expected (the middle right panel of Fig.~\ref{fig:effects_everthing_v_age}), since more massive stars evolve faster than less massive stars. We find no evidence that the observed correlation with planet properties is an effect of stellar mass rather than age (see previous Section, see also the top panels and the middle-left panel of Fig.~\ref{fig:effects_everthing_v_age}). 
We further look at age as a function of orbital period (the middle-left panel of Fig.~\ref{fig:effects_everthing_v_age}). While many planets in our sample orbit at short periods, others have periods of several days and no particular correlation as a function of age can be observed.
Overall we conclude that no particular sample bias can be identified that would lead to the observed effect. The most plausible bias would be an absence of low-mass planets orbiting young stars due to the challenge of RV observations. We find no evidence of this occurring, but also note that such an effect could only explain the absence of low-density planets around young stars, and not the similar absence of high-density planets around older stars.

%\begin{adjustbox}{center}
\begin{table*}
\begin{tabular}{m{0.1cm}  m{3.0cm}m{0.8cm}m{0.8cm}m{0.8cm}m{0.8cm}m{0.8cm}m{0.8cm}m{0.8cm}}
\hline
\hline 
{} &           Planet name &  R &  R$_{\rm{err}}$ &  M  & M$_{\rm{err}}$  &   Age &  Age$_{\rm{err,m}}$ &  Age$_{\rm{err,p}}$ \\
{} &           &  [R$_{\oplus}$] &  [R$_{\oplus}$] &  [M$_{\oplus}$] & [M$_{\oplus}$] &   [Gyr] &  [Gyr] &  [Gyr]\\
\\
\hline
\hline
1  &       Kepler-20 b &    1.80 &        0.08 &    9.46 &        1.32 &   3.04 &   2.33 &   3.76 \\
2  &  EPIC 249893012 b &    1.83 &        0.12 &    9.01 &        1.21 &   9.20 &   1.07 &   1.05 \\
3  &          K2-291 b &    1.52 &        0.11 &    6.35 &        1.18 &   3.46 &   2.37 &   4.65 \\
4  &         WASP-47 e &    1.74 &        0.07 &    6.86 &        0.71 &   6.43 &   3.52 &   4.23 \\
5  &  EPIC 220674823 b &    1.77 &        0.09 &    8.03 &        0.93 &  11.07 &   5.56 &   5.70 \\
6  &        HD 80653 b &    1.56 &        0.09 &    5.70 &        0.41 &   3.48 &   2.26 &   1.97 \\
7  &         CoRoT-7 b &    1.53 &        0.14 &    5.83 &        0.68 &   3.42 &   2.57 &   3.63 \\
1  &      Kepler-107 c &    1.50 &        0.10 &    9.81 &        2.03 &   5.06 &   2.36 &   2.54 \\
2  &      Kepler-107 b &    1.44 &        0.10 &    3.71 &        1.76 &   5.06 &   2.36 &   2.54 \\
10  &          K2-131 b &   1.43 &        0.09 &    7.63 &        1.31 &   6.52 &   4.34 &   7.22 \\
11 &         HD 3167 b &    1.64 &        0.15 &    4.92 &        0.39 &  10.62 &   6.21 &   5.22 \\
12 &          K2-111 b &    1.76 &        0.12 &    5.39 &        0.80 &  14.03 &   2.43 &   2.53 \\
13 &      Kepler-100 b &    1.31 &        0.06 &    5.92 &        2.50 &   6.72 &   0.93 &   0.83 \\
14 &         TOI-561 b &    1.35 &        0.07 &    2.19 &        0.23 &  12.19 &   6.43 &   4.30 \\
15 &      Kepler-323 c &    1.73 &        0.09 &    6.19 &        3.21 &  11.25 &   5.36 &   5.00 \\
16 &           K2-38 b &    1.52 &        0.10 &    7.68 &        1.21 &   4.59 &   3.06 &   3.92 \\
17 &        HD 20329 b &    1.67 &        0.10 &    7.31 &        0.85 &  13.50 &   4.62 &   3.39 \\
18  &       Kepler-21 b &   1.60 &        0.05 &    7.14 &        1.27 &   4.28 &   0.61 &   1.2  \\
19 &          55 Cnc e &    1.86 &        0.08 &    8.20 &        0.63 &   7.26 &   4.63 &   6.26 \\
20 &       Kepler-93 b &    1.48 &        0.03 &    4.64 &        0.55 &   6.21 &   1.41 &   1.83 \\
21 &       Kepler-10 b &    1.45 &        0.06 &    3.25 &        0.41 &   9.84 &   4.68 &   4.57 \\
22 &          KOI-94 b &    1.50 &        0.22 &    8.80 &        3.76 &   3.32 &   1.96 &   1.90 \\
23 &      Kepler-406 b &    1.32 &        0.06 &    6.22 &        1.33 &   1.76 &   1.21 &   2.81 \\
24 &        TOI-1444 b &    1.41 &        0.09 &    3.74 &        0.72 &   8.52 &   5.17 &   6.55 \\
25 &       HD 213885 b &    1.68 &        0.08 &    8.44 &        0.88 &   4.56 &   2.73 &   3.22 \\
26 &       HD 137496 b &    1.19 &        0.07 &    4.08 &        0.63 &   7.57 &   1.81 &   1.82 \\
\bottomrule
\end{tabular}
\caption{Parameters of small, rocky exoplanets used in Figures 2 and 3. From left to right the table lists planet name, radius, mass and age, and their associated errors. The quoted lower (Age$_{\rm{err,m}}$) and upper (Age$_{\rm{err,p}}$) age uncertainties are the 16th and 84th percentiles directly extracted from the Bayesian Posterior Distribution Function (PDF) by \basta. For each planet, the references for our adopted values of $\frac{R_p}{R_\star}$ and Radial Velocity K-amplitude are given in Table~\ref{tab:planets_observations}.}
\label{tab:planets_properties}
\end{table*}
%\end{adjustbox}

\section{The galactic chemical context}
\label{sec:discussion}
Galactic chemical evolution predicts a link between stellar chemical abundances and ages. Older stars are expected to be enhanced in alpha-elements like O, Mg, Si. Younger stars are more metal-rich and less alpha-enhanced, as a result of enrichment of the interstellar medium by type Ia supernovae. Observations have qualitatively confirmed this picture \citep{ciuca2021,tuccimaia2016,vv2022}.

Previous studies have predicted the link between the composition of rocky planets to the the chemical composition of their host stars \citep[e.g.,][]{santos2017}. This was observed, as a mass-period-metallicity relation \citep{Sousa2019}, and later  when the star iron-mass-fraction, defined as  [Fe / Fe + Si + Mg], was shown to correlate to a modelled planet iron-mass-fraction, and to the planet bulk density \citep{Adibekyan21}. Abundance trends in refractory and volatile elements of solar twins have also been suggested to be a tracer of the formation of Earth-like planets \citep{melendez2009,ramirez2009}. Further, it is well understood that abundance ratios themselves, such as C/O and Mg/Si, directly influence planet properties, such as the size of the planetary core \citep{Dorn2015,Unterborn2014,Unterborn2107}.
On the other hand, recent re-analysis of planet radius and mass measurements suggest that the link between stellar Mg and Si abundances with rocky planet composition is significantly weaker than previously thought \citep{brinkman24}.
\par
A potential link between stellar composition and rocky planet composition has separately been suggested, due to recent work on polluted white dwarfs. \citet{Hollands2021} found that old, lithium-rich white dwarf spectra showed signs of crust-dominated planet debris accretion. \citet{Elms2022} presented analysis of the oldest-known polluted white dwarf. With the use of future Gaia releases, a larger sample of polluted white dwarfs can be used to verify the trend we see in this work, by comparing the composition of pollutants from the spectra, and their ages, which are related to cooling timescales.
\par
Our results demonstrate for the first time, via homogeneous host star characterisation, that these stellar compositional links with rocky exoplanet compositions are also linked to age. It is well known that the elements considered to be influential on rocky planet formation in this way, are strong tracers of stellar age \citep[for example, see the Si/Age relation in][]{Snaith2014}.
Our results therefore support robust theoretical predictions and suggestions that galactic chemical evolution is the driving factor for the compositional diversity of rocky exoplanets \citep{oneill2020, cabral2023}.

We also note that the [Mg/Si] ratio itself varies with age \citep{Bedell2018}, at a different rate than the alpha elements individually, due to the enhanced production of Si in comparison to Mg, by type Ia supernovae. In simulations, this ratio has been shown to directly impact the mass of a planet at a given radius \citep{Unterborn2107}, in agreement with our results.
\par
Fig.~\ref{fig:sample_youngerplanets_higheriron} shows the correlation between stellar age and rocky planet composition. It may further be interpreted that younger, more iron-rich stars, are able to form planets of a greater compositional diversity. Older ($>8$~Gyr), less iron-rich stars may not have been able to produce planets with e.g. $>50 \ironmassfrac$. This is also seen in Fig.~\ref{fig:investigating_ironmassfraction_v_mass}, where stars with $M_{s} > 0.95$~\stellarmass do not appear to host any alpha-rich, less dense rocky planets with $\ironmassfrac<20$. This may be due to the fact that higher mass stars typically belong to a sample of stars that were born more recently. Due to chemical evolution, they are more iron-rich, and less alpha-enhanced. Thus, they may be more likely to be able to form a dense, iron-rich rocky planet due to this chemical abundance.
\par
Fig. \ref{fig:investigating_ironmassfraction_v_mass} also demonstrates that [Fe/H] does not play a significant role in the observed trend, as for each mass bin [Fe/H] is relatively scattered. This is discussed further in Section 5.1 of \cite{brinkman24}.
\par
In comparing the left and right panels of Fig. \ref{fig:investigating_ironmassfraction_v_mass}, we tentatively observe that lower-mass stars appear to have fewer of the highest-density planets. However, we caution this effect could be caused by the relatively small sample size and more data are required to investigate this further. Finally, in comparing the left and right panels of Fig.\ \ref{fig:investigating_ironmassfraction_v_mass}, we also see that there are fewer planets orbiting higher mass stars at higher ages in our sample. We expect this is a natural consequence of the faster evolution of higher mass stars, which are on average younger. While not unexpected, this complicates disentangling the effect of stellar mass and stellar age on planet properties, and a larger sample is required to fully investigate these effects.
\par
\section{Conclusions}
\label{sec:conclusions}
We find that planets which formed more recently (about $2-6$~Gyr old), around younger stars, have a higher Fe content than planets which formed earlier ($6-14$~Gyr). Our findings imply that the composition of rocky planets in our solar neighbourhood is dependent on when, in the history of our Galaxy, these planets formed.  We interpret this as an effect of galactic chemical evolution. More metal-rich stars born more recently, may be more likely to host denser rocky planets, due to the higher abundance of iron in their discs.
\par
Although beyond the scope of this work, the relationship between rocky planet density and age for systems outside of the limitations of this sample could be considered. This may include planets orbiting M and early-K dwarfs (with $\teff< 5000$~K), which could be aged using alternative techniques such as gyrochonology \cite[e.g.][]{Engle2023} or kinematics \cite[e.g.][]{Sagear2024}. This could also be extended to younger ($<2$~Gyr) systems, when mass measurements of the planets become available. Probing such data would allow for a more thorough analysis of the source(s) of the observed correlation.
\par
Furthermore, the PLATO mission \citep{PLATO} is predicted to significantly increase the sample size of known small, rocky exoplanets. Alongside planet discovery, its goals include homogeneous stellar characterisation, including the use of asteroseismology to provide precise ages for planet host stars, which should allow for more data to be added to plots such as Fig.~\ref{fig:sample_youngerplanets_higheriron}, potentially allowing for more in-depth investigation to the physical causes of the correlation.
These denser, more Fe-rich planetary cores are important for planetary processes such as plate tectonics \citep{Shah2022,Unterborn2022}, life-protecting magnetic dynamos \citep{Rickard2007} and affect the planetary surface gravity.
The results of this work suggest that age, in addition to stellar composition and mass, is an important parameter to consider when interpreting the composition of rocky planets, their formation, and their potential habitability. 
 %\clearpage

\section*{Acknowledgements}

A.W. would like to thank the Science and Technology Facilities Council (STFC) for funding support through a PhD studentship. 
V.V.E. and D.K. have been supported by UK's Science \& Technology Facilities Council through the STFC grants ST/W001136/1 and ST/S000216/1. 
D.H. acknowledges support from the Alfred P. Sloan Foundation, the National Aeronautics and Space Administration (80NSSC19K0597, 80NSSC21K0652), and the Australian Research Council (FT200100871).
D.K. was supported by MWGaiaDN, a Horizon Europe Marie Sk\l{}odowska-Curie Actions Doctoral Network funded under grant agreement no. 101072454 and also funded by UK Research and Innovation (EP/X031756/1).
A.S. has been supported by funding from the European Research Council (ERC) under the European Union’s Horizon 2020 research and innovation programme (CartographY; grant agreement ID 804752). Funding for the Stellar Astrophysics Centre was provided by The Danish National Research Foundation (Grant agreement No.~DNRF106). P.P. acknowledges funding from the European Research Council (ERC) under the European Union’s Horizon Europe research and innovation programme (grant agreement No 101076489)

\section*{Data Availability}
This research has made use of the NASA Exoplanet Archive, which is operated by the California Institute of Technology, under contract with the National Aeronautics and Space Administration under the Exoplanet Exploration Program. This work presents results from the European Space Agency (ESA) space mission Gaia. Gaia data are being processed by the Gaia Data Processing and Analysis Consortium (DPAC). Funding for the DPAC is provided by national institutions, in particular the institutions participating in the Gaia MultiLateral Agreement (MLA). The Gaia mission website is https://www.cosmos.esa.int/Gaia. The Gaia archive website is https://archives.esac.esa.int/Gaia. All data used for analyses in the paper are available in Tables 1, 2 and 3.
\\
\noindent
This study makes use of \texttt{numpy} (\url{https://numpy.org}), \texttt{scipy} (\url{https://scipy.org}), \texttt{matplotlib} \citep{hunter2007}, \texttt{astropy} \citep{astropy}, \texttt{astroquery} \citep{astroquery}, \texttt{BASTA} (\url{https://basta.readthedocs.io}), and \texttt{linmix} (\url{https://linmix.readthedocs.io}).

%%%%%%%%%%%%%%%%%%%% REFERENCES %%%%%%%%%%%%%%%%%%

% The best way to enter references is to use BibTeX:

\bibliographystyle{mnras}
\bibliography{example} % if your bibtex file is called example.bib

% Alternatively you could enter them by hand, like this:
% This method is tedious and prone to error if you have lots of references
%\begin{thebibliography}{99}
%\bibitem[\protect\citepauthoryear{Author}{2012}]{Author2012}
%Author A.~N., 2013, Journal of Improbable Astronomy, 1, 1
%\bibitem[\protect\citepauthoryear{Others}{2013}]{Others2013}
%Others S., 2012, Journal of Interesting Stuff, 17, 198
%\end{thebibliography}

%%%%%%%%%%%%%%%%%%%%%%%%%%%%%%%%%%%%%%%%%%%%%%%%%%

%%%%%%%%%%%%%%%%% APPENDICES %%%%%%%%%%%%%%%%%%%%%

%%%%%%%%%%%%%%%%%%%%%%%%%%%%%%%%%%%%%%%%%%%%%%%%%%

\appendix

\section{}

\subsection{Validation of Stellar Masses and Radii}
\begin{figure*}
\centering
\includegraphics[width=0.9\textwidth]{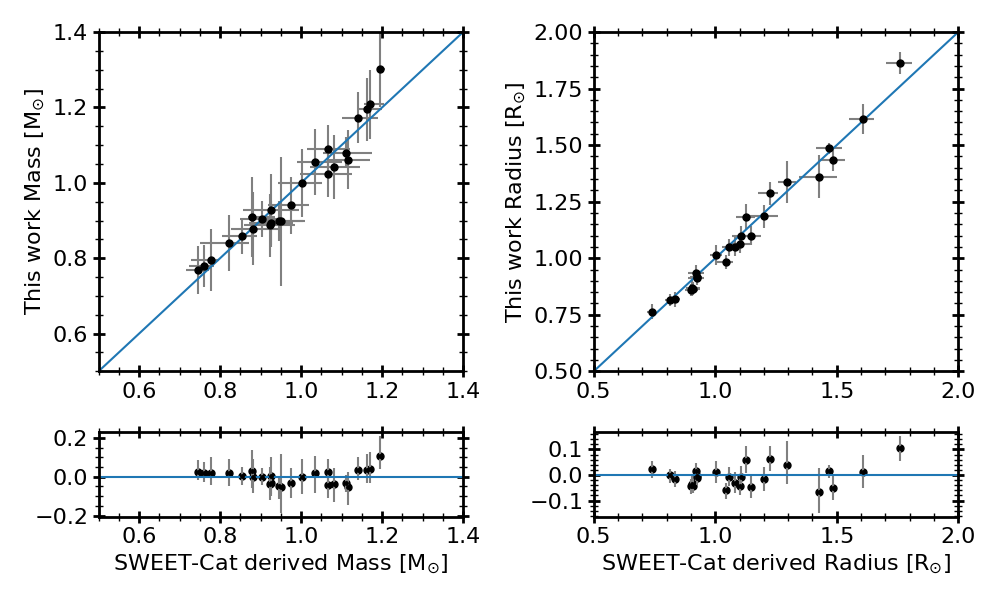}
\caption{Masses and radii for the 25 stars in our sample, as determined by using our \basta stellar properties inference method using to different sets of spectroscopic inputs: Gaia DR3 GSP-Spec (y-axis) and \sweetcat (x-axis). We see excellent agreement, with median residual scatters of 4.2$\%$ and 4.1$\%$.
}
\label{fig:validation_sweetcatcomparison}
\end{figure*}

We validate our stellar masses and radii by using input atmospheric parameters from the \sweetcat catalogue \citep{Santos2013,Sousa2021}. \sweetcat is a catalogue of spectroscopic parameters for exoplanet host stars, extracted from high-precision spectra and fitted to model atmospheres. These input parameters are not fully homogeneous: about $25\%$ of the \sweetcat sample has spectroscopic parameters extracted completely homogeneously. 
%This is to validate that the Gaia DR3 GSP-Spec parameters are appropriate for this method of stellar characterisation.
We use the same grid and formalism as described earlier in Section~\ref{sec:basta}, but replace the input spectroscopic parameters metallicity [M/H] and effective temperature \teff~ from GSP-Spec pipeline with ~\feh~ and ~\teff~ from \sweetcat. Fig.~\ref{fig:validation_sweetcatcomparison} shows radius and mass as determined by our original method and inputs as well as the results found using the same method but input parameters from \sweetcat instead. %on the x-axes, and on the y-axis we show the values determined by the SWEET-cat validation method.
We see excellent agreement with a median residual scatter of 4.2$\%$ and 4.1$\%$ respectively, validating that input parameters from Gaia DR3 GSP-Spec measurements are appropriate for this method of stellar characterisation. Due to their homogeneity, we adopt the Gaia DR3 GSP-Spec measurements for our final stellar characterisation.

\begin{figure*}
\centering
\includegraphics[width=0.9\textwidth]{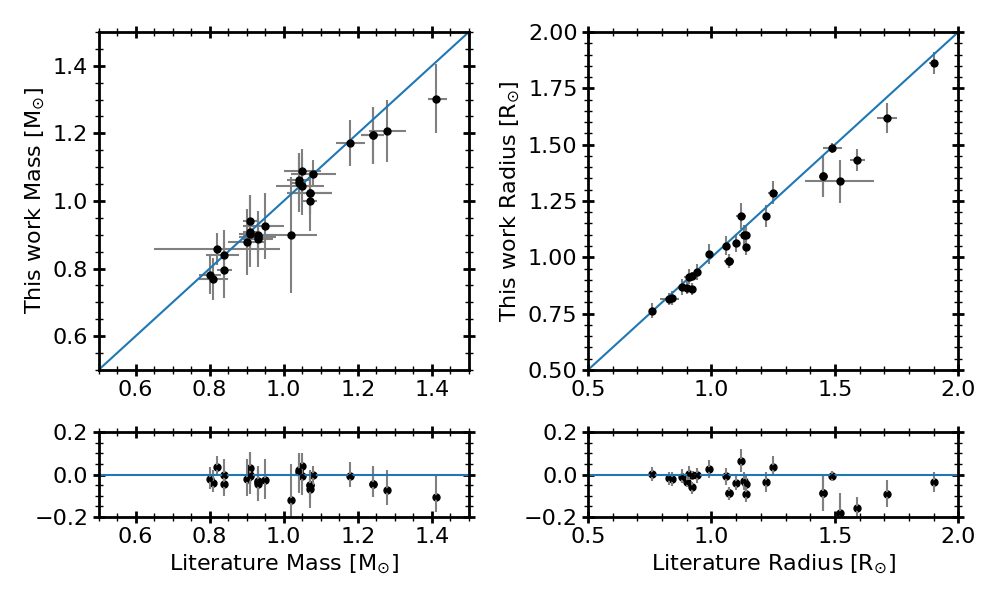}
\caption{Masses and radii for the 25 stars in our sample, as determined by using our \basta fitting method using Gaia DR3 GSP-Spec spectroscopic parameters (y-axis), plotted as a function of literature parameters. We see excellent agreement, with median residual scatter of 4.7 $\%$ and 3.8 $\%$.
}

\label{fig:validation_literaturecomparison}
\end{figure*}

We further compare the stellar masses and radii derived here to literature values. In Figure \ref{fig:validation_literaturecomparison}, we show the stellar masses and radii derived here, relative to the literature values currently adopted for those stars in the NASA exoplanet archive. Despite the heterogeneity of the comparison values, we again find excellent agreement, with a median residual scatter of 4.7$\%$ and 3.8$\%$.

 \subsection{Validation of Stellar Ages}
\label{sec:agevalidation}
%We further aim to validate our stellar ages. Age is one of the most challenging of the stellar parameters to estimate. We validate our ages by comparing subsets of our data to samples which have been studied with other methods. We compare to literature ages using carious techniques, and apply our method to a separate sample of stars with high-precision asteroseismic ages available.

We validate our stellar ages by comparing our approach to a range of other methods. In the next subsections, we discuss our ages in the context of asteroseismology, gyrochronology and chemical abundances, and kinematics.

\subsubsection{Asteroseismology}
\par

\begin{figure*}
\centering
    \includegraphics[width=0.45\textwidth]{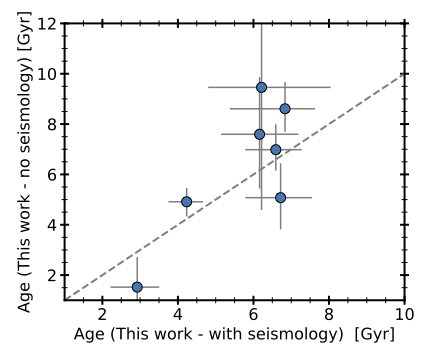}
    \includegraphics[width=0.45\textwidth]{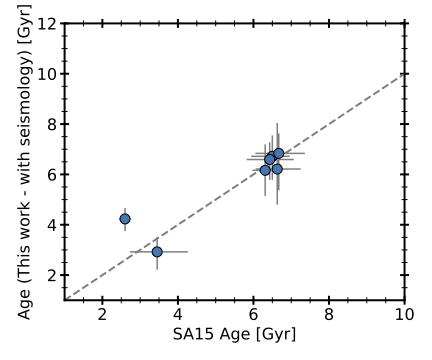}
    
    \caption{Left: Ages determined from \basta, for stars which exhibit asteroseismic oscillations visible in Kepler data, and reported in \citet{SA2015}. The y-axis shows ages determined without asteroseismic input to \basta, using the inputs described in the text. The x-axis shows the ages determined when the same input parameters are used, in addition to asteroseismic parameters. A 1:1 line is shown in grey. Right: The x-axis shows asteroseismic ages determined by \citet{SA2015}, for stars which overlap with our sample, plotted as a function of asteroseismic ages of the same stars in our analysis, on the y-axis.}
\label{fig:validation_asteroseismology_basta_seismic_v_noseismic}
\end{figure*}

First, we compared \basta output ages with and without including global asteroseismic parameters (\dnu~, \numax~).
The results are shown in Figure~\ref{fig:validation_asteroseismology_basta_seismic_v_noseismic} (left). %4.
We find a standard deviation for the absolute residuals of 0.9\,Gyr. We observe a median residual offset and standard deviation of $15 \pm 15\%$ and $40\%$, showing good agreement with our estimated uncertainties.

The gold standard for determining ages of single stars is asteroseismic modeling of individual oscillation frequencies. 
Figure~\ref{fig:validation_asteroseismology_basta_seismic_v_noseismic} (right) %5
shows our \basta output ages (including the asteroseismic \dnu~ and \numax~ as inputs) compared to ages determined from individual frequency modeling in \cite{SA2015}. We find a median absolute offset of 1.0\,Gyr and the standard deviation for the absolute residuals is 1.1\,Gyr.
For fractional residuals, we find an offset of $6 \pm 2\%$ and a residual scatter of $16\%$, again showing good agreement with our estimated uncertainties.
%\begin{figure}[h]
%\centering
%\includegraphics[width=0.9\textwidth]{figs/SA15_BASTA_age_casecase24_barbie_seismo.pdf}
%\caption{Ages from this work, using asteroseismic inputs, (y-axis) plotted as a function of the literature asteroseismology ages (x-axis, \citep{SA2015}), for host stars in our initial sample. he line of unity is shown as a dashed, grey line. We find a median offset and scatter of   $6.1 \pm 2.3\% $ and $16\%$}
%\label{fig:validation_asteroseismology_ages_basta_v_kages}
%\end{figure}

\begin{figure}
\centering
\includegraphics[width=0.48
\textwidth]{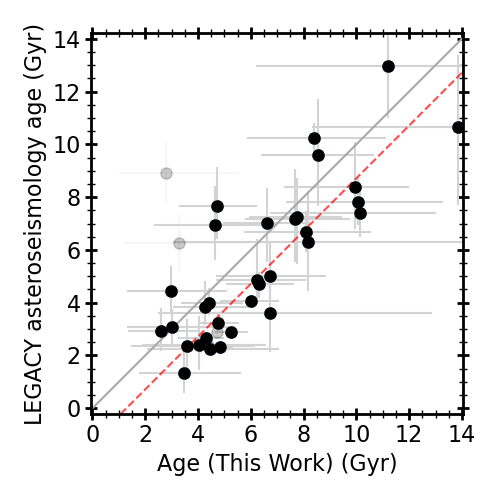}
\caption{36 Stars from the Kepler LEGACY sample with similar parameters to the stars in our sample. Ages determined with the method of this paper are plotted on the x-axis, against asteroseismic ages from \citet{aguirre2017} on the y-axis. The line of unity is shown as a grey line. The dashed red line shows this line of unity shifted to the median offset of $1.28$~Gyr. 4 stars with a difference in metallicity of 0.15 dex or greater between \citet{aguirre2017} and Gaia are in faint grey, with the remaining 36 in black.}
%The standard deviation of the absolute residuals is 2\,Gyr, which is smaller than the mean uncertainty of 3.2\,Gyr. The standard deviation of the fractional residuals is 47\%.}}
%There is a scatter of $47 \pm 7.8 \%$. The red line shows the 1:1 agreement line corrected for the median offset. Good agreement is shown between our isochrone-determined ages, and a precise asteroseismic sample.}}
\label{fig:validation_asteroseismology_ages_basta_v_keplerlegacy}
\end{figure}

To conduct a more extensive validation on a larger sample of stars, we apply our methods to a separate sample of stars. For this, we use the Kepler LEGACY sample, for which ages are precisely determined using asteroseismology \citep{lund2017,aguirre2017}. Specifically, stellar parameters in this study were determined from frequency analysis of high quality, long baseline (minimum 12 month) photometric time series data from the Kepler mission. We use the stars in this sample which are within the stellar parameter range of our sample ($5000$ < \teff~ < $6500$~\si{\kelvin}, $0.7<\stellarmass<1.4$~\si{\solarmass}). We extract the same Gaia parameters ($\varpi$, $G$, $G_\mathrm{BP}$, $G_\mathrm{RP}$), following the same data preparation steps, and perform fitting with \basta and identical grids of stellar models for this sample as was done with the main sample in this paper. Figure~\ref{fig:validation_asteroseismology_ages_basta_v_keplerlegacy} shows the ages derived by our method relative to the asteroseismic ages from \cite{aguirre2017}. We find a median absolute offset of 1.28 Gyr, a median residual scatter of $47$~$\%$, and the standard deviation for the absolute residuals is 2 Gyr.
These results are consistent with our median uncertainties of 3.2 Gyr and 46$\%$.

\subsubsection{Gyrochronology \& Chemical Abundances}

\begin{figure}
\centering
\includegraphics[width=0.48\textwidth]{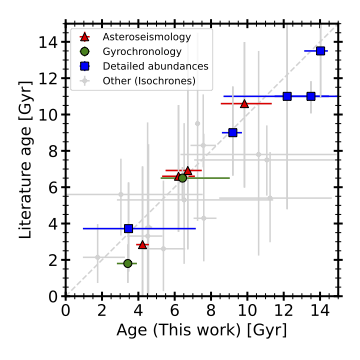}
\caption{Ages from this work, which were used in the final analysis, plotted as a function of the most recent literature ages, for host stars in our initial sample. The 1:1 fit is shown with the dashed grey line. Ages determined in the literature from gyrochronology are shown in green, those with well-constrained ages from measured \alphafe abundances are shown in blue, and those with asteroseismic literature values are shown in red.}
\label{fig:validation_asteroseismology_ages_basta_v_literature}
\end{figure}

As a further validation, we compare our ages to literature ages for these systems. Some were determined before the availability of Gaia DR3 parallaxes, and many use different stellar model tracks. Nevertheless, we compare the ages derived here to literature stellar ages, as shown in Figure~\ref{fig:validation_asteroseismology_ages_basta_v_literature}. Stars with ages calculated based on detailed chemical abundances, for example known low-iron hosts and stars with specific alpha element abundances reported, are highlighted in the blue squares, i.e., TOI-561, EPIC~249893012, K2-111, HD~20329, K2-291 \citep{brinkman23, Murgas22, hidalgo2020, Kosiarek2019,bourrier2022}. Stars with ages derived from age-rotation relations and measured rotation periods are shown in green, i.e., WASP-47, and CoRoT-7-b \citep{Almenara2016, john2022}. Stars with asteroseismic ages are shown in red, i.e., Kepler-21, Kepler-10, Kepler-93, Kepler-100, from \citep{SA2015}. All other stellar ages for this sample are shown in grey - these are primarily from isochrone ages from grids built with different models and differing input values on a star to star bases, making them difficult to compare. We see a good agreement for those stars where literature ages are available from asteroseismology, gyrochronology, or detailed stellar abundances. For all stars in this sample, including the grey points which are determined heterogeneously in the literature by isochrones, we find a median residual offset of $10.8 \%$ and a scatter of  $29.1 \%$. For the three categories of stars with precise ages, we find a median scatter of 1 Gyr. After correcting for this systematic effect, we find a median scatter of $12.9\%$, and the standard deviation for the absolute residuals is $0.97$~Gyr. These results are consistent with our median uncertainties of $40.1\%$ and $3.2$~Gyr.

\begin{figure}
\centering
\includegraphics[width=0.5\textwidth]{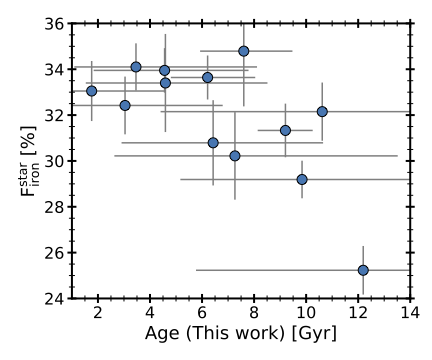}
\caption{Stellar iron-mass-fractions \starironmassfrac \citep{Adibekyan21} versus ages determined in this work. Stars with higher \starironmassfrac, thus more iron compared to alpha elements, are younger. This is expected from galactic chemical evolution, providing independent validation for our age determination. 
%These are the stars which are shown to host planets with higher iron mass fractions, and higher densities.
}\label{fig:validation_literature_ironmassfraction_v_our_ages}
\end{figure}

Finally, we compare our ages to the star iron-mass-fraction as estimated by \cite{Adibekyan21}, for those stars common to both studies. The result is shown in Figure~\ref{fig:validation_literature_ironmassfraction_v_our_ages}.
We observe that stars with enhanced value of \starironmassfrac are younger, whereas stars with enhanced alpha elements (and thus lower iron-mass-fraction) are older. This qualitatively confirms the expected picture from galactic chemical evolution. The lowest iron-fraction star in the sample is TOI-561, which was independently confirmed as an old star which is a member of the chemical Galactic thick disc \citep{weiss21,lacadelli21,brinkman23}.

\par
\par

\subsubsection{Kinematics}

\begin{figure}
\centering
\includegraphics[width=0.48
\textwidth]{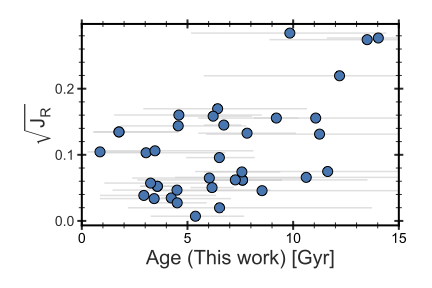
}
\caption{Dynamical radial actions ($\sqrt{J_R}$) versus ages from this work. We broadly observe that stars with older ages in our sample correspond to stars that have undergone a higher degree of secular orbital evolution.}
%This quantity are weak kinematic indicators of age. Although correlation is not strong, with a Pearson coefficient of 0.47, we verify that generally older ages in our sample correspond to stars that have undergone a higher degree of secular orbital evolution.}
\label{fig:validation_kinematics_jr_v_ages}
\end{figure}

We further compare our ages to dynamical radial actions ($J_R$), which provide a measure for the radial excursions of a star's orbit around the centre of the Milky Way. We calculate these using the \verb+`galpy'+ \footnote{\url{https://docs.galpy.org/en/v1.9.2/\#acknowledging-galpy}} Python package, using the \verb+MWPotential2014+ potential \citep{Bovy2015}. The radial actions are calculated according to
\begin{equation}
    J_i = \frac{1}{2\pi} \oint_{\mathrm{orbit}} P_i \,\mathrm{d}x_i
\end{equation}
 where $i$ = $(r, \theta, z)$.
Dynamical actions are a weak age indicator, due to the fact that orbits of stars born in the plane kinematically heat up over time due to interactions with substructure in the Galaxy \citep{Beane2018}. We plot age and normalized $\sqrt{J_R}$ of host stars in the rocky planet sample, in Figure~\ref{fig:validation_kinematics_jr_v_ages}. We observe a weak positive trend, in broad agreement with expected chrono-kinematic relations.

\subsubsection{Summary of Age Validation}
The tests above demonstrate that our ages derived using Gaia input parameters have relative accuracies and precision of $\approx$2\,Gyr ($\approx$ 30\% for the median age in our sample). Since our sample is homogeneous, our conclusions are very minimally impacted by relative age accuracy, and are primarily based on relative age precision. The precision achieved with our method is consistent with independent studies who have used Gaia parallax to derive ages \citep{Berger2018,Berger2023a} and have identified demographic trends of exoplanet properties with age \citep[e.g.,][]{Berger2020,Berger2023b,past2,past3,past4}. 
%, validating the statistical significance of the correlation in Figure~\ref{fig:sample_youngerplanets_higheriron}.

% Don't change these lines
\bsp	% typesetting comment
\label{lastpage}
\end{document}